\documentclass[sigconf,screen,nonacm]{acmart}

\AtBeginDocument{%
  \providecommand\BibTeX{{%
    \normalfont B\kern-0.5em{\scshape i\kern-0.25em b}\kern-0.8em\TeX}}}

\setcopyright{acmcopyright}
\copyrightyear{2021}
\acmYear{2021}
\acmDOI{10.1145/0000000.0000000}

\acmConference[Woodstock '18]{Woodstock '18: ACM Symposium on Neural
  Gaze Detection}{June 03--05, 2018}{Woodstock, NY}
\acmBooktitle{Woodstock '18: ACM Symposium on Neural Gaze Detection,
  June 03--05, 2018, Woodstock, NY}
\acmPrice{15.00}
\acmISBN{978-1-4503-XXXX-X/18/06}

\hypersetup{hidelinks=false,colorlinks=true,linkcolor=blue,urlcolor=blue,citecolor=blue,anchorcolor=blue}

\usepackage{url} 

\usepackage{tabularx} 
\usepackage{booktabs} 
\usepackage{stfloats}          

\usepackage[maxfloats=150]{morefloats}                                               
\usepackage{microtype}                                                               
\usepackage[utf8]{inputenc}   

\newcommand\Invisible[1]{                                                            
  \marginpar{\color{white}{\fontsize{.5}{.5}\selectfont #1 }}                        
}  

\newcommand{\Exclude}[1]{}

\definecolor{Gray95}{gray}{0.95}
\definecolor{forestgreen}{rgb}{0.13, 0.55, 0.13}

\newtheorem{theorem}{Theorem}
\newcommand{\code}[1] {\texttt{#1}}
\newcommand{\subs}[2] {#1\textsubscript{#2}}
\newcommand{\Self}[1] {\subs{Self}{#1}}
\newcommand{\remove}[1] {}

\newcommand{\AtFoot}[1]{\let\thefootnote\relax\footnotetext{{#1}}}

\usepackage{microtype}   
\usepackage{listings} 
\usepackage{graphicx} 

\makeatletter
\renewcommand*\verbatim@nolig@list{}
\makeatother

\usepackage[T1]{fontenc}
\usepackage[newfloat=true,frozencache=true]{minted}
\usepackage{changepage} 

\usepackage[iso,american,cleanlook]{isodate}                                         

\usepackage[export]{adjustbox} 

\usepackage{fancyhdr}                                                             
\makeatletter
\lst@Key{countblanklines}{true}[t]%
    {\lstKV@SetIf{#1}\lst@ifcountblanklines}

\lst@AddToHook{OnEmptyLine}{%
    \lst@ifnumberblanklines\else%
       \lst@ifcountblanklines\else%
         \advance\c@lstnumber-\@ne\relax%
       \fi%
    \fi}
\makeatother

\lstdefinestyle{numbers}
{numbers=left, stepnumber=1, numberstyle=\tiny, numbersep=10pt}
\lstdefinestyle{nonumbers}
{numbers=none}

\usepackage{tikz} 
\usepackage[inline]{enumitem}

\newcommand{\orcidicon}[1]{\href{https://orcid.org/#1}{\XeTeXLinkBox{\includegraphics[scale=0.06]{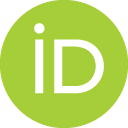}}}}

\begin{document}


\title{Hemlock : Compact and Scalable Mutual Exclusion} 

\author{Dave Dice \orcidicon{0000-0001-9164-7747}}
\email{david.dice@gmail.com}
\orcid{0000-0001-9164-7747}
\affiliation{\institution{University of Waterloo}\country{Canada}} 
\author{Alex Kogan \orcidicon{0000-0002-4419-4340}}
\email{alex.kogan@oracle.com}
\orcid{0000-0002-4419-4340}
\affiliation{\institution{Oracle Labs}\country{USA}} 
\renewcommand{\shortauthors}{Dice and Kogan}

\begin{abstract}
We present \textbf{Hemlock}, a novel mutual exclusion locking algorithm that is 
extremely compact, requiring just one word per thread plus one word per lock,
but which still provides \emph{local spinning} in most circumstances, 
high throughput under contention, and 
low latency in the uncontended case.  Hemlock is \emph{context-free} -- not requiring
any information to be passed from a lock operation to the corresponding unlock -- and FIFO. 
The performance of Hemlock is competitive with and often better than the best scalable spin locks.  
\AtFoot{This is an extended version of \cite{spaa21-Dice}} 
\end{abstract}




\maketitle
\thispagestyle{fancy} 
\fancyfoot[C]{\vspace{0.5cm} \today \hspace{1mm} \textbullet \hspace{1mm}} 


\section{Introduction} 

Locks often have a crucial impact on the performance of parallel software, 
hence they remain in the focus of intensive research.
Many locking algorithms have been proposed over the last several decades.
Ticket Locks \cite{focs79,cacm79-reed,tocs91-MellorCrummey} are simple and compact,
requiring just two words for each lock instance and no per-thread data.  
They perform well in the absence of contention, exhibiting low latency because
of short code paths.  Under contention, however, performance suffers \cite{EuroPar19-TWA} 
because all threads contending for a given lock will busy-wait on a central location, increasing
coherence costs.  For contended operation, so-called \emph{queue based} locks, such
as CLH\cite{craig-clh,ipps94-magnusson} and MCS\cite{ppopp91-Mellor-Crummey} provide relief
via \emph{local spinning} \cite{topc15-dice}.  For both CLH and MCS, arriving threads
enqueue an element (sometimes called a ``node'') onto the tail of a queue and then
busy-wait on a flag in either their own element (MCS) or the predecessor's element (CLH).  
Critically, at most one thread busy-waits on a given location at any one time, increasing
the rate at which ownership can be transferred from thread to thread relative to
techniques that use global spinning, such as Ticket Locks. 

Hemlock is inspired by CLH, and, like CLH, threads wait on a field associated with the predecessor.  
Hemlock, however, avoids the use of queue nodes, freeing the implementation from lifecycle 
concerns -- allocating, releasing, caching -- associated with that structure.  
The lock and unlock execution paths are extremely simple.  An uncontended lock operation
requires just an atomic SWAP (exchange) operation, and unlock just a compare-and-swap (\texttt{CAS}), 
which is the same as MCS.  Like Ticket Locks, MCS and CLH locks, 
Hemlock provides FIFO admission.  

Hemlock is compact,  requiring just one word per extant lock plus 
one word per thread, regardless of the number of locks held or waited upon.
Like MCS and CLH, the lock contains a pointer to the tail of the queue of threads waiting on
that lock, or \texttt{null} if the lock is not held. 
The thread at the head of the queue is the owner. 
In MCS the queue is implemented as an explicit linked list running from the head (owner) 
to the tail.  In CLH the queue is implicit and each thread waits on a field in
its predecessor's element.   CLH also requires that a lock in unlocked state be
provisioned with an empty queue element. When that lock is destroyed, the element must be recovered.
Hemlock avoids that requirement.  

\Invisible{CLH: pre-initialize or initialize on-demand} 
\Invisible{Hemlock is degenerate CLH} 
\Invisible{Oracle invention disclosure accession number : ORC2133687-US-NPR (Compact and Scalable Mutual Exclusion)} 

Instead of using queue nodes, Hemlock provisions each thread
with a singular \texttt{Grant} field where any immediate successor can busy-wait.  Normally
the \texttt{Grant} field -- which acts as a mailbox between a thread and its 
immediate successor on the queue -- is \texttt{null}, indicating empty.  During an unlock operation,  
where there are threads queued behind the owner, the outgoing owner
installs the address of the lock into its \texttt{Grant} field and then waits for
that field to return to \texttt{null}. 
The successor thread observes the lock address appear in its predecessor's \texttt{Grant} field, 
which indicates that ownership has transferred.  The successor then responds by clearing the \texttt{Grant} 
field, acknowledging receipt of ownership and allowing the \texttt{Grant} field of its 
predecessor to be reused in subsequent handover operations, and then finally enters the critical section.  

\Invisible{Simple contention; unary contention} 

Under simple contention, when a thread holds at most one contended lock at a time, 
Hemlock provides local spinning.  
But if we have one thread $T1$ that holds multiple contended locks, the immediate successors 
for each of the queues will busy-wait on $T1$'s \texttt{Grant} field.  
As multiple threads (via multiple locks) can be busy-waiting on $T1$'s \texttt{Grant} field,
$T1$ writes the address of the lock being released into its own \texttt{Grant} field to
disambiguate and allow the specific successor to determine that ownership has been conveyed.  
We note that simple contention is a common case for many applications. 
This is supported by the surveys of Cheng et al. \cite{spaa98-cheng} and 
O'Callahan et al. \cite{ppopp03-ocallahan} -- which found that it is rare for a thread to
hold multiple locks at a given time -- as well as by our profiling of LevelDB, described below.
This suggests that Hemlock would enjoy local spinning in many practical settings. 
\Invisible{Cheng report that it is uncommon, in the applications they surveyed, 
for a thread to hold multiple locks at a given time, suggesting that Hemlock would 
enjoy local spinning.}  

\Invisible{Disambiguate succession by writing lock address into \texttt{Grant}.} 

\Invisible{Potential names : hemlock; spinoff; tailspin; forespin; prospector; systolic; lockstep } 

\Invisible{
N-waiting where N is the number of distinct contended locks held by a thread.
Devolves to non-local spinning : multi-waiting
prefixes for X-local waiting : 
usually; mostly; frequently; commonly; generally; dominantly; 
semi; pseudo; oft; 
fere; ferme; consuete; pluries; multo; multum; praecipue; maxime; saepo; frequens; 
plerumque; plerusque; 
valde; fortiter; vehementer; 
} 

The advantages of avoiding queue nodes, however, does not end in reduced space and a simplified 
implementation that avoids node lifecycle concerns.  
Both CLH and MCS need to convey the address of the owner (head) node 
from the lock operation to the unlock operation.   The unlock operation
needs that node to find the successor, and to reclaim nodes from the queue
so that nodes may be recycled.  
While the locking API could be modified to accommodate this requirement,
it is inconvenient for the classic POSIX \texttt{pthread} locking interface, 
in which case the usual approach to support MCS is to add a field to the lock 
body that points to the head, allowing that value to be passed from the lock operation to the 
corresponding unlock operation.  
This new field is protected by the lock itself, but accesses to the field
execute within the effective critical section and may also induce additional coherence traffic.
Instead of an extra field in the 
lock body to convey the head -- the owner's element -- from the lock operation to
the corresponding unlock, implementations may also opt to use a per-thread associative map that relates 
the lock address to the owner's element.   
A lock algorithm or interface that does not need to pass information from the lock operator 
to unlock is said to be \emph{context-free} \cite{ppopp16-wang}. 
Hemlock is context-free and does not require the head pointer in the unlock operation,
simplifying the implementation.  

\Invisible{
@   Like CLH, threads in Hemlock wait on state in their predecessor's element in the queue.  
@   Invariant : while multiple threads might wait on Self->Grant,
    at most one thread at any time will be waiting for L to appear in Self->Grant.
@   Invariant : Except immediately after admission when a waiting thread has
    been granted a lock and needs to clear grant from L to 0,
    only the owner (Self) can write Self->Grant.
@   Potential extra interlock and delay in unlock()
@   Grant acts as a contended-for singleton mailbox
    Unlock must wait for previous unlock() to read and release mailbox
@   inform successor via Self->Grant
@   Thread waits on state in the predecessor like CLH
@   mailbox : empty vs full
@   Announce handover via Grant
    publish; announce; convey; grant;
@   take turns handing over ownership to successors on multiple locks
    must wait for ack and mailbox empty
@   chokepoint in unlock
@   During handover, outgoing/departing owner must wait
    for successor to observe L in mailbox and acknowledge by
    clearing mailbox back to empty null state.
@   Interlock stall executes outside CS and after handover
@   occupied; busy; full; vacant; empty; 
@   constellation; configuration; 
@   depict
@   Hemlock is arguably equivalent to CLH where CLH uses a singleton per-thread
    shared queue element
@   systolic
@   store; deposit; install
@   pipeline; 
@   synchronous; tightly coupled; interlock; dialog; conversation; back-and-forth; exchange; 
    request-response; reply; acknowledge; lock-step; 
@   deposit; store; place; install; 
@   Concern :
    Mailbox acts as a point of contention
    Mailbox induces coherence traffic
    However: a given thread can release only one lock at a time
} 

\Invisible{
It has some resemblance to MCS, CLH and ExactVM’s old MetaLock.
Each thread has a TLS Self variable and the thread structure contains
a Grant field.   A lock consists of just a tail pointer in the usual
MCS/CLH fashion.   Very much unlike CLH and MCS, there are no queue
nodes, but we have, at least under simple contention for a single lock,
strictly local spinning.   But if we have a thread T1 that holds multiple
locks, and there are threads waiting on those locks, then we do end up
with (kind of) global spinning on a field in T1’s thread structure as
the lead waiting thread for each of those locks spins will spin a common
field in T1.    Like CLH, we spin on a field associated with the predecessor.
The Grant field acts as a kind of mailbox to announce succession and
handover of ownership.   If it’s occupied, then unlock has to wait
for it to become empty, so arguably there’s an extra interlock where we
could stall if a thread unlocks L1 and L2 in quick succession and both
those locks are contended.   The fast uncontended path requires a SWAP
in lock and a \texttt{CAS} in unlock, so it’s on par with MCS in that respect,
although the paths are tighter.
}


\section{The Hemlock Algorithm}




\lstset{language=Python}
\lstset{frame=lines}
\lstset{basicstyle=\footnotesize\ttfamily} 
\lstset{commentstyle=\itshape\color{gray}} 
\lstset{commentstyle=\slshape\color{gray}} 
\lstset{commentstyle=\itshape\color{gray}} 
\lstset{keywordstyle=\color{forestgreen}\bfseries} 
\lstset{backgroundcolor=\color{Gray95}} 

\begin{figure*}[t]
\noindent\begin{minipage}[t]{0.45\linewidth}
\lstset{caption={Simplified Pseudo-code for Hemlock}}
\lstset{label={Listing:hemlock0}}
\begin{lstlisting}[language=Python,mathescape=true,escapechar=\%]
class Thread :
  atomic<Lock *> Grant = null 

class Lock :
  atomic<Thread *> Tail = null 

def %\underline{Lock}% (Lock * L) :
  assert Self$\rightarrow$Grant $==$ null        
  ## Enqueue self at tail of implicit queue
  auto pred = swap (&L$\rightarrow$Tail, Self) 
  if pred $\neq$ null :
     ## Contention : must wait
     while pred$\rightarrow$Grant $\neq$ L : Pause
     pred$\rightarrow$Grant = null
  assert L$\rightarrow$Tail $\neq$ null
  
def %\underline{Unlock}% (Lock * L) :
  assert Self$\rightarrow$Grant $==$ null
  auto v = cas (&L$\rightarrow$Tail, Self, null)
  assert v $\neq$ null
  if v $\neq$ Self :
     ## one or more waiters exist -- convey ownership to successor
     Self$\rightarrow$Grant = L 
     while Self$\rightarrow$Grant $\neq$ null : Pause

\end{lstlisting}
\end{minipage}\hfill%
\begin{minipage}[t]{0.45\linewidth}
\lstset{caption={Hemlock with CTR Optimization}}
\lstset{label={Listing:hemlockCTR}}
\begin{lstlisting}[language=Python,mathescape=true,escapechar=\%]
class Thread :
  atomic<Lock *> Grant = null 

class Lock :
  atomic<Thread *> Tail = null 

def %\underline{Lock}% (Lock * L) :
  assert Self$\rightarrow$Grant $==$ null
  auto pred = swap (&L$\rightarrow$Tail, Self) 
  if pred $\neq$ null :
     while cas(&pred$\rightarrow$Grant, L, null) $\neq$ L : Pause
  
def %\underline{Unlock}% (Lock * L) :  
  assert Self$\rightarrow$Grant $==$ null
  auto v = cas (&L$\rightarrow$Tail, Self, null)
  if v $\neq$ Self :
    Self$\rightarrow$Grant = L 
    while FetchAdd(&Self$\rightarrow$Grant, 0) $\neq$ null : Pause

\end{lstlisting}
\end{minipage}%
\end{figure*}

We start by describing a simplified version of the Hemlock algorithm, with pseudo-code 
provided in Listing-\ref{Listing:hemlock0}.  
In Section \ref{CTR}  we describe a key performance optimization, and the 
pseudo-code for that optimized Hemlock algorithm is given in Listing-\ref{Listing:hemlockCTR}.  

\texttt{Self} refers to a thread-local structure containing the thread's \texttt{Grant} field.  
Threads arrive in the lock operator at line 8 
and atomically swap their own address into the lock's \texttt{Tail} field, obtaining
the previous tail value, constructing the implicit FIFO queue. 
If the \texttt{Tail} field was \texttt{null}, then the caller acquired the lock without contention and
may immediately enter the critical section.  Otherwise the thread waits for the lock's address to appear in
the predecessor's \texttt{Grant} field, signalling succession, at which point the thread
restores the predecessor's \texttt{Grant} field to \texttt{null} (empty) 
indicating the field can be reused for subsequent unlock operations by the predecessor.  
The thread has been granted ownership by its predecessor and may enter the critical section.  
Clearing the \texttt{Grant} field, above, is the only circumstance in which one thread may
store into another thread's \texttt{Grant} field.  
Threads in the queue hold the address of their immediate predecessor, obtained as the return
value from the \texttt{SWAP} operation, but do not know the identity of their successor, if any. 

In the unlock operator, at line 16, threads initially use an
atomic compare-and-swap (\texttt{CAS}) operation to try to swing the lock's \texttt{Tail} field
from the address of their own thread, \texttt{Self}, back to \texttt{null}, which represents \emph{unlocked}. 
If the \texttt{CAS} was successful then there were no waiting threads and the 
lock was released by the \texttt{CAS}.
Otherwise waiters exist and the thread then writes the \emph{address} of the lock $L$ into its own \texttt{Grant}, 
alerting the waiting successor and passing ownership.  Finally, the outgoing thread waits for that successor
to acknowledge the transfer and restore the \texttt{Grant} field back to empty, indicating the field 
be reused for future locking operations.  Waiting for the mailbox to return to \texttt{null} happens outside
the critical section, after the thread has conveyed ownership.  

In Hemlock, transfer of ownership in unlock is \textbf{address-based}, where the outgoing 
owner writes the lock address into its own grant field, whereas under MCS and CLH owner transfer 
is via a boolean written into a queue element monitored by its immediate successor.

Threads that attempt to release a lock that they do not hold will stall indefinitely at Line 21, waiting  
for an acknowledgement that will never arrive.  This property makes it easy to identify and debug the offending
thread and unlock operation. 

The \texttt{assert} statements in the listings are not necessary for correct operation, but
serve to document invariants that may be useful in understanding the algorithm.  

\Invisible{Proximate and immediate cause; root cause} 

MCS and Hemlock allow trivial implementations of the \texttt{TryLock} operations -- 
using \code{CAS} instead of \code{SWAP} -- whereas Ticket Locks and CLH do not. 
An uncontended lock acquisition requires an atomic \texttt{SWAP} for MCS, CLH and Hemlock and 
an atomic \texttt{fetch-and-add} for Ticket Locks.  An uncontended unlock -- no waiters -- requires
an atomic \texttt{CAS} for MCS and Hemlock, and simple stores for CLH and Ticket Locks while
a contended unlock, which passes ownership to a waiter, requires a store for MCS, CLH and Ticket Locks.


In Listing-\ref{Listing:hemlock0} line 21, threads in the unlock operator must wait for the successor to 
acknowledge receipt of ownership, indicating the unlocking thread's \texttt{Grant} mailbox is again available 
for communication in subsequent locking operations.  
That is, the recipient needs to take positive action and respond before the previous owner can 
return from the unlock operator.  
While this phase of waiting occurs outside and after the transfer of ownership -- 
crucially, \emph{not} within the effective critical section or on critical path-- 
such waiting may still impede the progress and latency
of the thread that invoked unlock.  Specifically, we have tightly coupled back-and-forth synchronous communication, 
where the thread executing unlock stores into its \texttt{Grant} field and then waits for a response from the successor,
while the successor, running in the lock operator, waits for the transfer indication (line 11) and then responds to
the unlocking thread and acknowledges by restoring \texttt{Grant} to \texttt{null} (line 12).  
The unlock operator must await a positive reply from the successor in order to safely reuse the \texttt{Grant} field 
for subsequent operations.  That is, the algorithm must not start an unlock operation until the previous
contended unlock has completed, and the successor has emptied the mailbox.  
We note that MCS, in the unlock operator, must also wait for 
the successor executing in the lock operator to establish the back-link that allows the owner to
reach the successor.  That is, both MCS and Hemlock have wait loops in the contended \texttt{unlock} path
where threads may need to wait for the arriving successor to become visible to the current owner,
and as such, neither unlock operator is wait-free.  
Compared to MCS and CLH, the only additional burden imposed by Hemlock that falls inside the critical path 
is the clearing of the predecessor’s \texttt{Grant} field by the recipient (Line 12),
which is implemented as a single store.  

To mitigate the performance concern described above, we could optimize Hemlock 
to defer and shift the waiting-for-response phase (Listing-\ref{Listing:hemlock0} line 21) to 
the prologue of subsequent lock and unlock operations, allowing more useful overlap and 
concurrency between the successor, which clears the \texttt{Grant} field, 
and the thread which performed the unlock operation.  The thread that called \texttt{unlock} 
may enter its critical section earlier, before the successor clears \texttt{Grant}.  
Ultimately, however, we opted to forgo this particular optimization in our 
implementation as it provided little observable performance benefit.  
While the \texttt{Grant} mailbox field might appear to be a source of contention and
to potentially induce additional coherence traffic, a given thread can release only one 
lock at a time, mitigating that concern. 

\Invisible{Overlap; pipelining;}

\subsection{Optimization: Coherence Traffic Reduction} 
\label{CTR} 

The synchronous back-and-forth communication pattern where a thread waits for ownership
and then clears the \texttt{Grant} field (Listing-\ref{Listing:hemlock0} Lines 11-12) is
inefficient on platforms that use MESI or MESIF ``single writer'' cache coherence protocols \cite{MESIF,Hennessy}. 
Specifically, in \texttt{unlock} when the owner stores the lock address into its 
\texttt{Grant} field (Line 20), it drives the cache line underlying \texttt{Grant} into
\emph{M}-state (modified) in its local cache. Subsequent polling by the successor (Line 11) 
results in a \emph{coherence miss} that 
will pull the line back into the successor's cache in $S$-state (shared).  
The successor will then observe the waited-for lock address and proceed to clear \texttt{Grant} 
(Line 12) forcing an upgrade from $S$ to $M$ state in the successor's cache and 
invaliding the line from the cache of the previous owner, adding a delay in the 
critical path where ownership is conveyed to the successor. 

We avoid the upgrade coherence transaction by polling with \texttt{CAS} 
(Listing-\ref{Listing:hemlockCTR} Line 9) instead of using simple loads, so, 
once the hand-over is accomplished and the successor observes the lock address, 
the line is already in $M$-state in the successor's local cache.  
We refer to this technique as the \emph{Coherence Traffic Reduction (CTR)} optimization. 
 
As an alternative to busy-waiting with \texttt{CAS}, we can achieve equivalent performance
by using an atomic \texttt{fetch-\allowbreak{}and-\allowbreak{}add} of $0$ -- 
implemented via \texttt{LOCK:XADD on x86} --
on \texttt{Grant} as a \emph{read-with-intent-to-write} primitive, and, after observing the waited-for 
lock address to appear in \texttt{Grant}, issuing a normal store to clear \texttt{Grant}. 
That is, we simply replace the \texttt{load} instruction in the traditional busy-wait loop with
\texttt{fetch-and-add} of $0$.  
Busy-waiting with an atomic read-modify-write operator, such as 
\texttt{CAS},\texttt{SWAP} or \texttt{fetch-and-add}, is typically considered a performance 
anti-pattern. For instance, Anderson\cite{tpds90-Anderson} observed that 
\texttt{test-and-test-and-set} locks are superior to crude \texttt{test-and-set} 
locks when there are multiple waiters.  
But in our case with the 1-to-1 communication protocol used on the \texttt{Grant} 
field in Hemlock, busy-waiting via read-modify-write atomic operations provides a performance benefit. 
Because of the simple communication pattern, back-off in the busy-waiting
loop is not useful.  

We also apply CTR in the unlock opertor at Listing-\ref{Listing:hemlockCTR} Line 15 as 
we expect the \texttt{Grant} field will be written by that same thread in subsequent 
\texttt{unlock} operations.  

Related approaches to coherence-optimized waiting have been described \cite{DiceMoir}.
Using \texttt{MONITOR-MWAIT}\cite{cluster13-akkan} to wait for invalidation, instead of 
waiting for a value, has promise, but the facility is not yet available in user-mode on Intel processors
\footnote{Future Intel processors may support user-mode \texttt{umonitor} and 
\texttt{umwait} instructions \cite{umwait}.  We hope to use those instructions in future Hemlock experiments}. 
\texttt{MWAIT} may confer additional benefits, as it avoids a classic busy-wait loop and thus
avoids branch mispredictions in the critical path to exit the loop when ownership 
has transferred \cite{DiceMoir-SpinLoopExit}. 
In addition, depending on the implementation, \texttt{MWAIT} may be more ``polite'' with respect
to yielding pipeline resources, potentially allowing other threads, including the lock owner,
to execute faster by reducing competition for shared resources. 
We might also busy-wait via hardware transactional memory, where invalidation of lines
in a processor's read-set or write-set will cause an abort,
serving as a hint to the waiting thread.  In addition, other techniques to hold the line
in $M$-state are possible, such as issuing stores to a dummy variable that abuts the
\texttt{Grant} field but which resides on the same cache line.  Using the \texttt{prefetchw} 
prefetch-for-write advisory ``hint'' instruction would appear workable but yielded no performance 
improvement in our experiments
\footnote{We plan on experimenting with non-temporal stores and new \texttt{CLDEMOTE} instruction, 
with the intention that the writer can immediately expunge the written-to-line from its cache, avoiding
subsequent coherence traffic when the reader loads that line.}. 

\Invisible{Polite waiting : altruism that ultimately benefits the benefactor} 
\Invisible{Confer; afford; yield; admit} 

The CTR optimization is specific to the shared memory communication pattern used
in Hemlock, and is not directly applicable to other lock algorithms.  

All Hemlock performance data reported herein uses the CTR optimization unless
otherwise noted.  We note that the relative benefit of CTR is retained on 
single-socket non-NUMA Intel sysetms. 

\Invisible{Benefit of CTR is not specific to Intel NUMA systems.} 
\Invisible{We note tha CTR is equally effective on non-NUMA single-socket Intel systems.} 

In Section \ref{Section:OffCore} we show the impact of the CTR optimization
on coherence traffic.

\subsection{Example Configuration} 
\label{ExampleConfiguration} 

\begin{figure}[h!]
\includegraphics[width=8cm,frame]{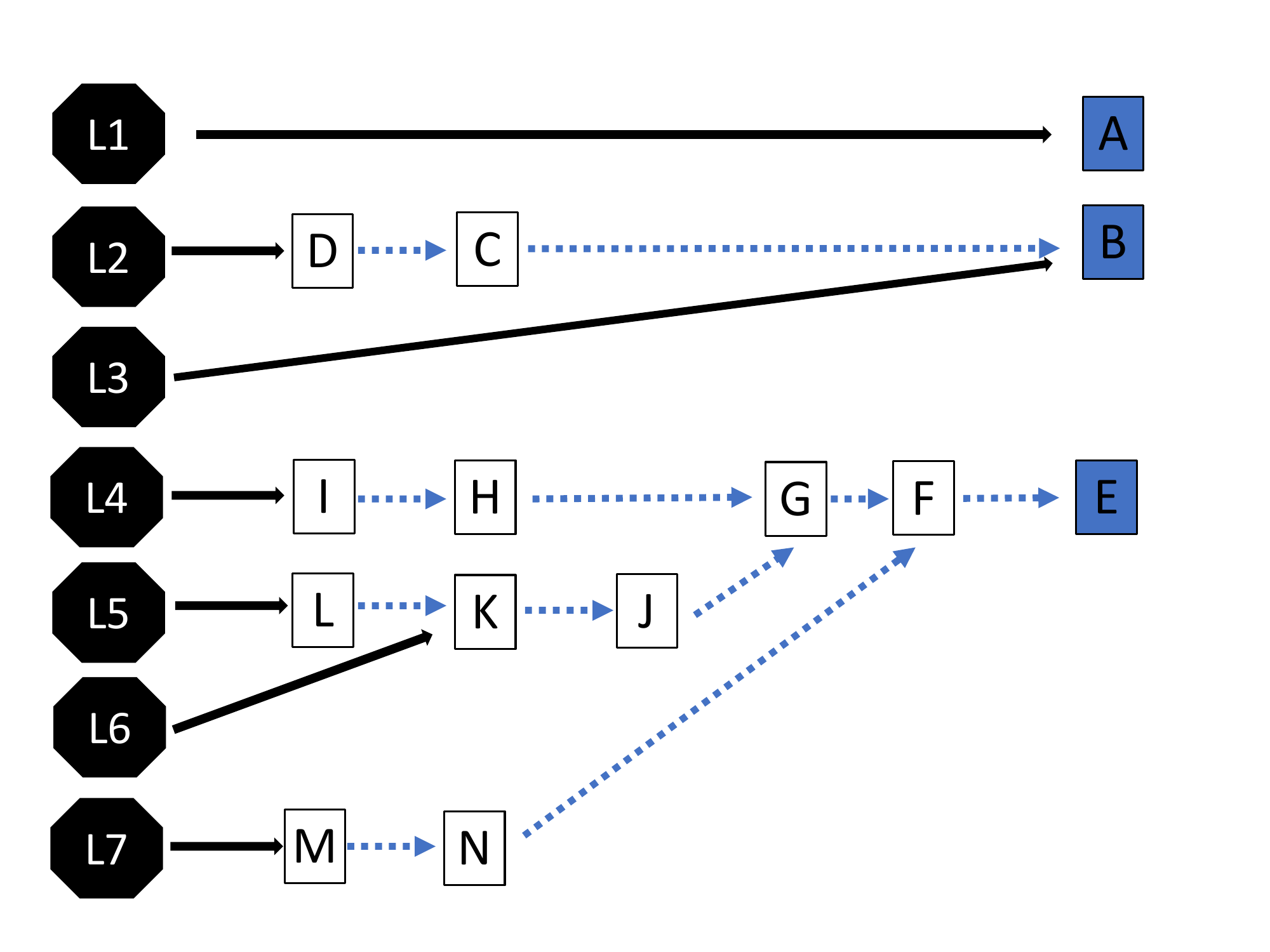} 
\vspace{-8pt}      
\caption{Object Graph in Hemlock}
\label{Figure:Hemlock-graphB}
\end{figure}

Figure-\ref{Figure:Hemlock-graphB} shows an example configuration of a set of 
threads and locks in Hemlock. 
$L1-L7$ depict locks while $A-N$ represent threads. 
Solid arrows reflect the lock's explicit \texttt{Tail} pointer, 
which points to the most recently arrived thread -- the tail of the lock's queue.  
Dashed arrows, which appear between threads, refer to a thread's immediate predecessor in the 
implicit queue associated with a lock.  
The address of the immediate predecessor is obtained via the atomic \texttt{SWAP} executed when threads arrive.
The dashed edge can be thought of as the \emph{busy-waits-on} relation and are not 
physical links in memory that could be traversed.  
In the example, $A$ holds $L1$, $B$ holds $L2$ and $L3$ while $E$ holds $L4$, $L5$ and $L7$. 
$K$ holds $L6$ but also waits to acquire $L5$.  
$A$, $B$ and $E$ execute in their critical sections.  
while all the other threads are stalled waiting for locks. 
The queue of waiting threads for $L2$ is $C$ (the immediate successor) followed by $D$. 
$D$'s predecessor is $C$, and, equivalently, $C$'s successor is $D$. 
Thread $D$ busy-waits on $C$'s \texttt{Grant} field and $C$ busy-waits on 
$B$'s \texttt{Grant} field. 

Threads $H$ and $J$ both busy-wait on $G$'s \texttt{Grant} field.  
In simple locking scenarios Hemlock provides local waiting, but when the dashed
lines form junctions (elements with in-degree greater than one) in the waits-on  
directed graph, we find non-local spinning, or \emph{multi-waiting}.   
Similarly, in our contrived example, $N$ and $G$ both wait on $F$.  
While our design admits inter-lock performance interference, arising from multiple
threads spinning on one \texttt{Grant} variable, as is the case for $G$ and $F$, 
above, we believe this case to be rare and not of consequence for common applications. 
(For comparison, CLH and MCS does not allow the concurrent sharing of queue elements, and
thus provides local spinning, whereas Hemlock has a shared singleton queue element 
-- effectively the \texttt{Grant} field -- that can be subject to being busy-waited
upon by multiple threads).  
Crucially, if we have a set of coordinating threads where each thread 
acquires only one lock at a time, then they will enjoy local spinning.
Non-local spinning can occur only when threads hold multiple locks.  
Specifically, the worst-case number of threads that could be busy-waiting on a given thread $T$'s
\texttt{Grant} field is $M$ where $M$ is the number of locks held simultaneously by $T$.  
We note that common usage patterns such as hand-over-hand ``coupled'' locking do not 
result in multi-waiting.  

\Invisible{
Non-local spinning occurs only when thread $T$ holds $L1$ and $T$ also
holds or waits for $L2$ and $T$ has waiting predecessors on both $L1$ and $L2$. 
\emph{multi-waiting} degree. 
} 

When $E$ ultimately unlocks $L4$, $E$
installs a pointer to $L4$ into its \texttt{Grant} field. Thread $F$ observes that store,
assumes ownership, clears $E$'s \texttt{Grant} field back to empty (null) and enters the critical section.  
When $F$ then unlocks $L4$, it deposits $L4$'s address into its own \texttt{Grant} field. 
Threads $G$ and $N$ both monitor $F$'s \texttt{Grant} field, with $G$ waiting for $L4$ to appear
and $N$ waiting for $L7$ to appear.  Both threads observe the update of $F$'s \texttt{Grant} 
field, but $N$ ignores the change while $G$ notices the value now matches $L4$, the lock
that $G$ is waiting on, which indicates that $E$ has passed ownership of $L4$ to $G$.  
$G$ clears $F$'s \texttt{Grant} field, indicating that $F$ can reuse that field for 
subsequent operations, and enters the critical section.  

We note that holding multiple locks does not itself impose a performance penalty,
while holding multiple locks when contention (and waiting) is involved will result in reduced
local spinning and a consequent reduction in performance because of increased
lock handover latency.

\Invisible{ 
Threads $A$, $E$ and $G$ are 
monitoring that field and observe the update. $E$ and $G$ see but ignore the change while
$A$ notices the value now matches $L1$, the lock it is waiting on, which indicates that $H$ has 
passed ownership of $L1$ to $A$. $A$ then clears $H$'s \texttt{Grant} field to empty 
(\texttt{null}) and then enters the critical section protected by $L1$.  
} 

\Invisible{
Characterization : 
@  Space usage and requirements 
@  atomic counts for contended-uncontended lock-unlock 
@  RMR counts for contended-uncontended lock-unlock
@  RMR complexity
}

\subsection{Space Requirements}  

\begin{table} [h]
\centering
\begin{tabular}{lccccc}
\toprule
\multicolumn{1}{l}{} &
\multicolumn{5}{c}{Space}      \\
\cmidrule(lr){2-6}

&
\multicolumn{1}{c}{Lock} &
\multicolumn{1}{c}{Held} &
\multicolumn{1}{c}{Wait} &
\multicolumn{1}{c}{Thread} & 
\multicolumn{1}{c}{Init}\\
\midrule

MCS           & 2      & $E$ & $E$  & 0 &           \\
CLH           & 2+$E$  & 0   & $E$  & 0 & $\bullet$ \\
Ticket Locks  & 2      & 0   & 0    & 0 &           \\
Hemlock       & 1      & 0   & 0    & 1 &           \\

\midrule[\heavyrulewidth]
\bottomrule
\end{tabular}%
\caption{Space Usage}\label{Space}
\end{table}

Table-\ref{Space} characterizes the space utilization of MCS, CLH, Ticket Locks, and Hemlock.
The values in the \emph{Lock} column reflect the size of the lock body in words.  For MCS and 
CLH we assume that the implementation stores the head of the chain -- reflecting the current 
owner --  in an additional field in the lock body, and thus the lock consists of 
\texttt{head} and \texttt{tail} fields, requiring 2 words in total. 
\emph{E} represents the size of a queue element. 
CLH requires the lock to be preinitialized with a so-called \emph{dummy} element before use.  
When the lock is ultimately destroyed, the current dummy element must be recovered. 
The \emph{Held} field
indicates the space cost for each held lock and similarly, the \emph{Wait} field indicates
the cost in space of waiting for a lock.  The \emph{Thread} column reflects per-thread state that 
must be reserved for locking.  For Hemlock, this is the \texttt{Grant} field.
A single word suffices, although to avoid false sharing we opted to sequester the 
\texttt{Grant} field as the sole occupant of a cache line.  In our implementation we
also elected to align and pad the MCS and CLH queue nodes to reduce false sharing
and to provide a fair comparison, raising the size of $E$ to a cache line.  
\emph{Init} indicates if the lock requires non-trivial constructors and destructors.  
CLH, for instance, requires that the current dummy node be released when a lock is destroyed.  

Taking MCS as an example, lets say lock
$L$ is owned by thread $T1$ while threads $T2$ and $T3$ wait to acquire $L$.  The lock body for $L$
requires 2 words and the MCS chain consists of elements $E1$ $\Rightarrow$ $E2$ $\Rightarrow$ $E3$ 
where $E1$, $E2$ and $E3$ are associated with and enqueued by $T1$, $T2$ and $T3$ respectively.  
$L$'s \texttt{head} field
points to $E1$, the owner, and the \texttt{tail} field points to $E3$.  
The space consumed in this configuration is 2 words for $L$ itself plus $3*E$ 
for the queue elements.  In comparision, Hemlock consumes one word for $L$ and and 3 words
of thread-local state for the \texttt{Grant} fields.  

\Invisible{Hemlock is frugal and parsimonious in space.} 

In MCS, when a thread acquires a lock, it contributes an element to the associated queue, and
when that element reaches the head of the queue, the thread becomes the owner.  In the 
subsequent \texttt{unlock} operation, the thread extracts and reclaims that same element from the queue.
In CLH, a thread contributes an element but, and once it has acquired the lock, recovers
a different element from the queue -- elements \emph{migrate} between locks and threads.  
The MCS and CLH \texttt{unlock} operators require dependent loads and indirection  
to locate queue nodes, while Hemlock avoids these overheads. 
In MCS, if the \texttt{unlock} operation is known to execute in the same stack frame
as the \texttt{lock} operation, the queue element may be allocated on stack.  
This is not the case for CLH.  
As previously noted, Hemlock avoids elements and their management.  

The K42 \cite{K42,Scott2013} variation of MCS can recover the queue element before returning from \texttt{lock}
whereas classic MCS recovers the queue element in \texttt{unlock}.  
That is, under K42, a queue element is needed only while waiting but not while the lock is held,
and as such, queue elements can always be allocated on stack, if desired. 
While appealing, the paths are much more complex and touch more cache lines than the classic
version, impacting performance. 

If a lock site is well-balanced -- with the lock and corresponding unlock operators lexically scoped and
executing in the same stack frame 
\footnote{For example Java \texttt{synchronized} blocks and methods; C++ \texttt{std::lock\_guard} 
and \texttt{std::scoped\_lock}; or locking constructs that allow the critical section to be 
expressed as a lambda} 
-- a Hemlock 
implementation can opt to use an on-stack \texttt{Grant} field instead of the thread-local \texttt{Grant} 
field accessed via \texttt{Self}.  This optimization, which
can be applied on an ad-hoc site-by-site basis, also acts to reduce multi-waiting on the thread-local
\texttt{Grant} field.  

\Invisible{trade-off between multi-waiting and space; allows heterogenous mixed usage} 

\section{Correctness Proofs} 
\Invisible{Safety-exclusion and liveness} 
\Invisible{Nothing bad ever happens; something good eventually happens;} 

In this section, we argue that the Hemlock algorithm is a correct implementation of a mutual exclusion lock 
with the FIFO admission and so-called \emph{fere-local} \footnote{mostly or frequently local} spinning properties.
We define those properties more formally below, but first we note that we consider the standard model 
of shared memory~\cite{HW90} with basic atomic \emph{read} and \emph{write} operations as well as 
more advanced atomic \emph{SWAP}, \emph{CAS} and \emph{FAA} operations. 
We presume atomic operators with the usual semantics.

\Invisible{
The \texttt{SWAP} operation receives two arguments, address and new value, and atomically
reads the old value at the given address, writes the given new value and returns the old value. 
The \texttt{CAS} (compare-and-swap) operation receives three arguments, address, old value and new value, and atomically
reads the value at the given address, compares it to the given old value, and, if equal, writes the given new value.
We say in this case that the \texttt{CAS} is successful.
Otherwise, if the current value is different from the given old value, \texttt{CAS} does not do any change, 
and we say it is unsuccessful.
We assume the \texttt{CAS} instruction returns the current value it has read.
We note that given the return value, one can identify whether \texttt{CAS} has been successful by 
comparing that value to the old one used in the invocation of \texttt{CAS}.
Finally, the \texttt{FAA} (fetch-and-increment) instruction, which is required only for the 
optimized version of our algorithm,
receives two arguments, address and delta, and atomically
reads the old value at the given address, adds the delta (which can be any integer number), and writes the result back.
The \texttt{FAA} operation returns the old value, before the increment.
We note that while most existing hardware architectures support CAS, some do not support SWAP or FAA.
Where such support is not available, those instructions can be easily emulated using CAS.
}

\Invisible{Door-way vs door-step}  

Multiple threads perform execution steps, where at each step a thread may perform local computation or execute one
of the atomic operations on the shared memory. 
We assume threads use the Hemlock algorithm to protect access to one or more critical sections, i.e.,
specially marked blocks of code that must be executed by at most one thread at a time.
Our arguments are formulated for the simplified version of the algorithm given in Listing-\ref{Listing:hemlock0}, 
and as such, all line references in this section are w.r.t.\ Listing-\ref{Listing:hemlock0}.  
Yet, we note that the correctness arguments apply, albeit with minor modifications, to the 
optimized version in Listing-\ref{Listing:hemlockCTR}.  

We call Lines 5--13 the \emph{entry} code and Lines 14--21 the \emph{exit} code.
Each thread cycles between the entry code (where it is trying to get into the critical section), 
critical section code, exit code (where it is cleaning up to allow other threads to execute their critical sections) 
and the so-called remainder section, where it executes code that does not belong to any of the 
other three sections~\cite{Lynch96}.
We assume the order in which threads take their execution steps is unknown, yet no thread 
ceases execution in the entry, exit or critical sections.
In other words, if a thread $T$ is in any of those code sections at time $t$, it is guaranteed that, 
eventually, at some time $t^\prime>t$, $T$ would perform its next execution step.
We also assume that each thread executes a finite number of steps in the critical section.

We refer to Line 8 as the \emph{entry doorstep} of the entry code and Line 20 as 
the \emph{exit doorstep} of the exit code for lock $L$.
We say that a thread is \emph{spinning on a word W} if its next execution step is reading from a 
shared memory location $W$ inside the while-loop (e.g., in Line 11 in Listing-\ref{Listing:hemlock0}). 
We say a lock $L$ is \emph{associated} with a thread $T$ if $T$ has executed the entry doorstep for $L$, 
but has not completed the exit code for $L$.  We prove the following properties for Hemlock algorithm 
defined with respect to any instance of a lock $L$.

\begin{itemize}
\item \textbf{Mutual exclusion:} At any point in time, at most one thread is in the critical section.
\item \textbf{Lockout freedom:} Any thread that starts executing the entry code eventually completes the exit code.
\item \textbf{FIFO:} Threads enter the critical section in the order in which they execute the entry doorstep.
\item \textbf{Fere-local spinning:} At any point in time, the number of spinning threads on the same word is bounded by 
the maximum number of locks associated with any thread at that time.
\end{itemize}

We note that lockout-freedom is a stronger property than the more common deadlock-freedom property~\cite{Lynch96}. 
Also, we note that if a thread never executes entry code of one lock inside the critical section of another 
(i.e., each thread has at most one associated lock),
the fere-local spinning implies \emph{local} spinning, i.e., each spinning thread reads a different word $W$.
Furthermore, the fere-local spinning is a dynamic property, e.g., the bound at time $t$ does not depend 
on the maximum number of locks associated with a thread prior to $t$.

We start with an auxiliary lemma. 
We denote the \code{Self} variable that (contains the \code{Grant} field and) belongs to a thread $T_i$ as 
\code{\Self{i}}.

\begin{lemma}
For any lock $L$, if \code{L$\rightarrow$Tail} is \code{null}, there is no thread that executed the entry doorstep, 
but has not executed the exit doorstep.
In particular, there is no thread in the critical section protected by $L$.
\label{lemma:tail-null}
\end{lemma}

\begin{proof}
The claim trivially holds initially at the beginning of the execution when \code{L$\rightarrow$Tail} is \code{null}.

Let $T_j$ be the first thread for which the claim does not hold.
That is, $T_j$ is the first thread for which \code{SWAP} in Line 8 returns \code{null}, yet 
there is a thread $T_k$ that has executed that \code{SWAP} before $T_j$ but has not executed \code{CAS} 
in Line 16 yet. 

Let $T_i$ be the last thread that set \code{L$\rightarrow$Tail} to \code{null} before $T_j$ ($T_i$ might be the same 
thread as $T_k$ or a different one).  From the inspection of the code, once \code{L$\rightarrow$Tail} is set to 
a non-\code{null} value in the entry doorstep, it can revert to \code{null} only by a successful \code{CAS} 
in Line 16.  For \code{CAS} in Line 16 to be successful, it has to be executed by the last thread that 
executed \code{SWAP} in Line 8.
In other words, if $T_i$ executes a successful \code{CAS} in Line 16 at time $t_1$ and it executed the 
corresponding \code{SWAP} at time $t_0$ ($t_0 < t_1$), 
no other thread executed \code{SWAP} at time $t_0 < t < t_1$. 

Consider when $T_k$ executes the \code{SWAP} instruction in Line 8 w.r.t. $[t_0, t_1]$.
Case 1: $T_k$ executes \code{SWAP} at time $t_2 < t_0$.
Let $T_l$ be the next thread that executes \code{SWAP} after $T_k$ ($T_l$ can be the same as $T_i$, or a different thread).
Since \code{L$\rightarrow$Tail} contains \code{\Self{k}}, $T_k$ will enter the while-loop in Line 11.
It will exit the loop only when \code{\Self{k}$\rightarrow$Grant} changes to $L$, which can happen, according to 
the code, only in Line 20 when $T_k$ executes the exit code.
By induction on the number of threads that executed \code{SWAP} between $t_2$ and $t_0$, $T_i$ can execute 
the successful \code{CAS} in Line 16 only after $T_k$ executes the exit code. 
This means that when $T_i$ executes \code{SWAP} in Line 8, $T_k$ has executed \code{CAS} 
in Line 16 -- a contradiction.

Case 2: $T_k$ executes SWAP at time $t_2 > t_1$.
This means that when $T_j$ executes \code{SWAP} in Line 8, \code{L$\rightarrow$Tail} contains 
either \code{\Self{k}} or \code{\Self{l}} for some other thread $T_l$ that executes SWAP after $T_k$ -- 
a contradiction to the fact that $T_j$'s \code{SWAP} returned \code{null}.
\end{proof}

With this lemma, we prove the correctness property for Hemlock.

\begin{theorem}
The Hemlock algorithm provides mutual exclusion.
\end{theorem}

\begin{proof}
By way of contradiction, assume $T_i$ and $T_j$ are simultaneously in the critical section protected by the same lock $L$.
Let $t_i$ and $t_j$ be the points in time when $T_i$ and $T_j$ executed Line 8 for the last time, respectively.
Without loss of generality, assume $t_i > t_j$.
Consider the value returned by \code{SWAP} in Line 8 when executed by thread $T_i$.
If the returned value is \code{null}, by Lemma~\ref{lemma:tail-null} $T_j$ must have executed its \code{CAS} 
instruction in Line 16 before $t_i$.
Hence, $T_i$ will execute the critical section after $T_j$ has completed its own -- a contradiction.

If the returned value is \code{\Self{k} $\neq$ null} (for some thread $T_k$ that might be 
the same as $T_j$ or a different one), 
let $T_l$ be the thread that executes \code{SWAP} in Line 8 right after $T_j$ and before $T_j$ executes 
\code{CAS} in Line 16.
$T_l$ might be the same as $T_i$ or a different thread.
$T_l$ waits in Line 11 for \code{\Self{j}$\rightarrow$Grant} to become $L$.
\code{\Self{j}$\rightarrow$Grant} can only change to L (from \code{null}) in Line 20 by thread $T_j$.
When this happens, $T_j$ is outside of the critical section.
By induction on the number of threads that execute \code{SWAP} in $(t_j, t_i]$, when $T_i$ finds 
\code{\Self{j}$\rightarrow$Grant} to be \code{null} in Line 11
and subsequently enters the critical section, $T_j$ is outside of the critical section -- a contradiction.
\end{proof}

Next, we prove the progress property for Hemlock. We do so by showing first that a thread 
cannot get stuck in the exit code, i.e., the exit-code is lockout-free.

\begin{lemma}
Every thread $T_i$ exiting the critical section eventually completes the exit code.
\label{lemma:exit}
\end{lemma}

\begin{proof}
The only place in the exit code where a thread $T_i$ may iterate indefinitely is the while-loop in Line 21.
In the following, we argue that $T_i$ either completes the exit code without reaching Line 21, 
or eventually breaks out of the loop in Line 21.

Let $t_0$ be the time $T_i$ executes the last \code{SWAP} instruction in Line 8 before entering the critical section 
and $t_1$ be the time it executes the \code{CAS} instruction in Line 16 when it starts the exit code.
Consider the following two cases. 
Case 1: no thread executes \code{SWAP} in $[t_0, t_1]$.
In this case, \code{L$\rightarrow$Tail} contains \code{\Self{i}} and the \code{CAS} instruction in Line 16 is successful.
Therefore, \code{CAS} returns \code{\Self{i}}, and $T_i$ can complete the exit code by a constant number 
of its steps by skipping Lines 18--21.

Case 2: at least one thread executes \code{SWAP} in $[t_0, t_1]$.
Let $T_k$ be the first such thread.
Thus, \code{CAS} in Line 16 is not successful, and it returns \code{\Self{j}}, for some $j \neq i$ (perhaps $j = k$).
(We note that, by Lemma~\ref{lemma:tail-null}, \code{CAS} cannot return \code{null}.) 
Therefore, $T_i$ reaches the while-loop in Line 21, after storing $L$ into \code{\Self{i}$\rightarrow$Grant} in Line 20.
Consider the execution steps of thread $T_k$ after its \code{SWAP} instruction.
The \code{SWAP} instruction returns \code{\Self{i}}, and so $T_k$ reaches the while-loop in Line 11.
After $T_i$ executes Line 20, eventually $T_k$ reads $L$ from \code{\Self{i}$\rightarrow$Grant} and breaks out 
of the while-loop in Line 11.
Next, it executes Line 12, storing \code{null} into \code{\Self{i}$\rightarrow$Grant}.
Finally, $T_i$ eventually reads \code{null} in Line 21 and breaks out of the while-loop.
\end{proof}

Next, we show that a thread cannot get stuck in the entry code either, but first we prove a 
simple auxiliary lemma.

\begin{lemma}
The \code{SWAP} instruction in Line 8 executed by thread $T_i$ never returns \code{\Self{i}}.
\label{lemma:swap-result}
\end{lemma}

\begin{proof}
From code inspection, only thread $T_i$ can write \code{\Self{i}} into \code{L$\rightarrow$Tail}.
Thus, the claim holds until $T_i$ executes \code{SWAP} at least for the second time.

Let $T_i$ execute \code{SWAP} in Line 8 for the $k$-th time, $k \ge 2$, at time $t_k$.
Consider the previous, $k$-$1$-th execution of \code{SWAP} by $T_i$, at time $t_{k-1}$.
From code inspection, $T_i$ has to execute \code{CAS} in Line 16 at time $t_{k-1} < t < t_k$.
If \code{CAS} is successful, $T_i$ changes the value of \code{L$\rightarrow$Tail} to \code{null}, and thus $k$-th 
\code{SWAP} will return \code{null} or \code{\Self{j}} for $j \neq i$.
If \code{CAS} is unsuccessful, there has been (at least one) another thread $T_j$, $j \neq i$, 
that performed \code{SWAP} in Line 8 at time $t_{k-1} < t^\prime < t$.
Thus, $k$-th \code{SWAP} at time $t_k > t^\prime$ will return \code{\Self{j}}, or \code{\Self{k}} 
(for $k \neq i,j$) or \code{null}, but not \code{\Self{i}}.
\end{proof}

\begin{lemma}
Every thread $T_i$ starting the entry code eventually enters the critical section.
\label{lemma:entry}
\end{lemma}

\begin{proof}
The only place in the entry code where a thread $T_i$ may iterate indefinitely is the while-loop in Line 11.
In the following, we argue that $T_i$ either completes the entry code without reaching Line 11, 
or eventually breaks out of the loop in Line 11.

Consider the following two cases w.r.t.\ to the value returned by \code{SWAP} executed by 
thread $T_i$ in the entry code at time $t_i$.
Case 1: \code{SWAP} returns \code{null}. 
In this case, $T_i$ can complete the entry code by a constant number of its steps by skipping Lines 9--12.

Case 2: SWAP returns \code{\Self{j}}.
Thus, $T_i$ reaches the while-loop in Line 11 and waits until \code{\Self{j}$\rightarrow$Grant} contains $L$.
From Lemma~\ref{lemma:swap-result}, we know that $j \neq i$.
Consider the state of thread $T_j$ w.r.t.\ to the value returned by \code{SWAP} executed by 
thread $T_j$ in Line 8 at time $t_j < t_i$.
If $T_j$’s \code{SWAP} returned \code{null}, $T_j$ will complete the entry code, end eventually 
reach Line 20 in the exit code.
Otherwise, $T_j$’s \code{SWAP} returned \code{\Self{k}}.
If \code{\Self{k}} is equal to \code{\Self{i}}, then $T_i$ executed (another) \code{SWAP} 
at time $t_i^\prime < t_j$.
This means that $T_i$ has executed the exit code in the interval $(t_i^\prime, t_i)$, and in 
particular, has executed Line 20 in the interval $(t_j, t_i)$.
Therefore, $T_j$ will break out of the while-loop in Line 11*, enter the critical section, and 
eventually execute Line 20, allowing $T_i$ to complete its entry code.
If \code{\Self{k}} is not equal to \code{\Self{i}}, consider whether at time $t_j$, $T_k$ has 
completed the while-loop in Line 11 
(including by skipping that while-loop entirely by evaluating the condition in Line 9 to false).
If so, $T_k$ will complete the entry code, end eventually reach Line 20 in the exit code, 
letting $T_j$ and, eventually, $T_i$ to break out of the while-loop in Line 11*.
Otherwise, $T_k$ is waiting in Line 11. 
(We note that there is a third possibility that $T_k$ has executed \code{SWAP}, but has not evaluated 
the condition in Line 9 yet, or has evaluated it to true, 
but has not started the while-loop in Line 11. 
We treat it as one of the first two possibilities, according to whether or not $T_k$ eventually 
waits in the while-loop in Line 11).

In the case $T_k$ is waiting in Line 11, we consider recursively the state of $T_k$ w.r.t.\ 
to the value returned by its \code{SWAP}, and any thread $T_k$ is waiting for in Line 11.
Since the number of threads is bounded, there have to be two threads, $T_a$ and $T_b$, s.t.\ $T_a$’s 
\code{SWAP} returns \code{\Self{b}} and 
$T_b$’s \code{SWAP} returns either (a) \code{null}, or (b) \code{\Self{c}} for $T_c$ in 
$\{T_i, T_j, T_k, …, T_a\}$ or (c) \code{\Self{c}} for $T_c$ that has completed the 
while-loop in Line 11. .
Following the similar reasoning as above, we conclude that $T_b$ eventually executes Line 20. 
in its exit code, and allows $T_a$ to break our of the waiting loop in Line 11. .
By induction on the number of threads in the set $\{T_i, T_j, T_k, …, T_a\}$, we conclude that, eventually, 
$T_i$ completes the while-loop in Line 11 and enters the critical section.
\end{proof}

\begin{theorem}
The Hemlock algorithm is lockout-free.
\end{theorem}

\begin{proof}
This follows directly from Lemma~\ref{lemma:exit} and~\ref{lemma:entry}, and the assumption 
that a thread completes its critical section in a finite number of its execution steps.
\end{proof}

Next, we prove that threads enter the critical section in the FIFO order w.r.t.\ their execution of the entry doorstep.
In the following lemma, we show that when two threads execute the entry doorstep one after the other, the latter thread
cannot ``skip'' over the former and enter the critical section first.

\begin{lemma}
Let $T_i$ be the next thread that executes the entry doorstep after $T_j$.
Then $T_i$ enters the critical section after $T_j$.
\label{lemma:doorstep}
\end{lemma}

\begin{proof}
First, we note that the claim trivially holds if $i=j$. 
This is because $T_i$ may execute another entry doorstep only after (entering and) exiting the critical section.

Next, we consider two cases.
Case 1: $T_i$’s execution of the \code{SWAP} instruction in the entry doorstep returns \code{null}. 
This can only happen if $T_j$ performs \code{CAS} in Line 16 before $T_i$ executes the \code{SWAP} instruction.
This means, however, that $T_j$ has completed its critical section, and the claim holds.
Case 2: $T_i$’s execution of the \code{SWAP} instruction in the entry doorstep returns \code{\Self{j}}.  
Then $T_i$ will proceed to Line 11, and wait until \code{\Self{j}$\rightarrow$Grant} changes to $L$.
This can only happen when $T_j$ reaches Line 20, which means that, once again, $T_j$ has completed 
its critical section, and the claim holds.
\end{proof}

\begin{theorem}
The Hemlock algorithm has the FIFO property.
\end{theorem}

\begin{proof}
By way of contradiction, assume there is a thread $T_i$ that executes the entry doorstep after a 
thread $T_j$, but enters the critical section before $T_j$.
Without loss of generality, let $T_i$ be the first such thread in the execution of the algorithm.
Let $T_k$ be the thread that executes the entry doorstep right before $T_i$ ($T_k$ might be the 
same thread as $T_j$ or a different one).
By the way we chose $T_i$, $T_k$ has not entered the critical section when $T_i$ does.
This is a contradiction to Lemma~\ref{lemma:doorstep}.
\end{proof}

We are left to prove the last stated property of Hemlock, namely the fere-local spinning.
Again, we start with an auxiliary lemma.

\begin{lemma}
For every lock $L$ and thread $T_i$, there is at most one thread $T_j$ waiting in Line 11 for 
\code{\Self{i}$\rightarrow$Grant} to become $L$.
\label{lemma:one-waiter}
\end{lemma}

\begin{proof}
Consider thread $T_j$ waiting in Line 11 for \code{\Self{i}$\rightarrow$Grant} to become $L$.
To reach Line 11, $T_j$ executed Line 8, where \code{SWAP} returned \code{\Self{i}}.
This, in turn, means that $T_i$ has also executed Line 8 (before $T_j$ did).
This is because Line 8 is the only place where \code{\Self{k}} can be written into \code{L$\rightarrow$Tail}, 
for any thread $T_k$.

Assume by way of contradiction that another thread $T_k$ is also waiting in Line 11 
for \code{\Self{i}$\rightarrow$Grant} to become $L$.
Let $t_j$ and $t_k$ be the points in time when $T_j$ and $T_k$ executed Line 8 for the last time, respectively.
From the atomicity of \code{SWAP}, $t_j \neq t_k$.
Assume without loss of generality that $t_j < t_k$.
Let $t_i$ be the time $T_i$ executed the \code{SWAP} for the last time before $t_j$.
From the above, $t_i < t_j < t_k$.

From the inspection of the code, the only way for $T_k$ to write \code{\Self{i}} in Line 8 into \code{pred}
is for $T_i$ to execute Line 8 right before $T_k$ does.
That is, there has to be a point in time $t_j < t_i^\prime < t_k$ in which $T_i$ executed Line 8 again.
This means that during $(t_i, t_i^\prime)$, $T_i$ has completed the entry code, its critical section, 
and the exit code (and started executing another entry code).
When executing the exit code, $T_i$ performed \code{CAS} in Line 16*.
If this \code{CAS} is successful, this means that it takes place before $t_j$ (since the 
value of \code{L$\rightarrow$Tail} remains unchanged), 
and so $T_j$ would not read \code{\Self{i}} into \code{pred} in Line 8 at $t_j$.
Thus, this \code{CAS} has to fail, i.e., return a value different from \code{\Self{i}}.

Thus, $T_i$ has to execute Lines 20--21, and in particular, wait until its \code{Grant} field contains \code{null}.
This happens before $t_i^\prime$ and hence before $t_k$, therefore $T_j$ is the only thread at this point
that waits in Line 11 for \code{\Self{i}$\rightarrow$Grant} to become $L$. 
Since $T_i$ completes its exit point (and executes \code{SWAP} at $t_i^\prime$), 
it must have exited the while-loop in Line 21 before $t_i^\prime$.
This can happen only if $T_j$ has executed Line 12 after $t_j$ and before $t_i^\prime$.
Thus, $T_j$ no longer waits in Line 11 for \code{\Self{i}$\rightarrow$Grant} to become $L$ when $T_k$ starts 
to wait there -- a contradiction.
\end{proof}

Note that as explained in Section-\ref{ExampleConfiguration}, there might be multiple threads 
spinning on the word \code{\Self{i}$\rightarrow$Grant} in Line 11, each for a different lock $L$.
However, as we argue in the lemma above, there might be only one thread \emph{per any given lock} 
that waits for the value of \code{\Self{i}$\rightarrow$Grant} to change.

\begin{theorem}
The Hemlock algorithm has the fere-local spinning property.
\end{theorem}

\begin{proof}
Assume thread $T_i$ has $k$ associated locks at the given point in time. 
By inspecting the code, threads can spin on a word only in Lines 11 or 21.
By Lemma~\ref{lemma:one-waiter}, there might be at most $k$ threads spinning on 
\code{\Self{i}$\rightarrow$Grant} in Line 11,
one for each of the $k$ locks associated with $T_i$.
(We note that by the definition of the associated locks, a thread $T_j$ cannot spin 
on \code{\Self{i}$\rightarrow$Grant} and wait until it contains a value of a lock that is not associated with $T_i$.)
At the same time, only $T_i$ can spin on \code{\Self{i}$\rightarrow$Grant} in Line 21, and it does so after 
writing $L$ into \code{\Self{i}$\rightarrow$Grant} in Line 20.
This means that when $T_i$ starts spinning on \code{\Self{i}$\rightarrow$Grant}, another thread $T_j$ 
stops spinning on \code{\Self{i}$\rightarrow$Grant} in Line 11. 
We note that it can be easily shown that such $T_j$ exists.
Thus, at any given point in time, the number of threads spinning on \code{\Self{i}$\rightarrow$Grant} is bounded by $k$.
\end{proof}

\section{Related Work} 

While mutual exclusion remains an active research topic
\cite{ppopp17-Ramalhete} \cite{craig-clh} \cite{ppopp91-Mellor-Crummey} 
\cite{EuroPar19-TWA} \cite{icdcn20-jayanti} 
\cite{opodis17-dvir} 
\cite{EuroSys19-CNA} \cite{arxiv-CNA} 
\cite{eurosys17-dice}
\cite{dice2020fissile} 
\cite{oopsla99-agesen} 
\cite{topc15-dice} 
\cite{Scott2013} 
we focus on locks closely related to our design. 

Simple test-and-set or polite test-and-test-and-set \cite{Scott2013} locks 
are compact and exhibit excellent latency for uncontended operations, but fail 
to scale and may allow unfairness and even indefinite starvation.  Ticket Locks are compact
and FIFO and also have excellent latency for uncontended operations but they also 
fail to scale because of global spinning, although some variations attempt
to overcome this obstacle, at the cost of increased space
\cite{EuroPar19-TWA,spaa11-dice,Ticket-AWN}.  
For instance Anderson’s array-based queueing lock \cite{dc03-Anderson,tpds90-Anderson} 
is based on Ticket Locks but provides local spinning.  It employs a waiting 
array for each lock instance, sized to ensure there is at least one array element 
for each potentially waiting thread, yielding a potentially large footprint. 
The maximum number of participating threads must be known in advance when
initializing the lock.

Queue-based locks such as MCS or CLH are FIFO and provide local spinning and
are thus more scalable.  
MCS is used in the linux kernel for the low-level ``qspinlock'' construct
\cite{linux-locks,Long13,LWN2014}.  Modern extensions of MCS edit the queue
order to make the lock \emph{NUMA-Aware}\cite{EuroSys19-CNA}. 
MCS readily allows editing and re-ordering of the queue of waiting threads,
\cite{markatos,eurosys17-dice,EuroSys19-CNA} whereas editing the
chain is more difficult under Hemlock.  

Hemlock does not provide constant remote memory reference (RMR) complexity 
\cite{opodis17-dvir}. Similar to MCS, Hemlock lacks a wait-free unlock operation, whereas
the unlock operator for CLH and Tickets is wait-free.  Unlike MCS, Hemlock requires
active synchronous back-and-forth communication in the unlock path between the outgoing
thread and its successor.  

Dvir's algorithm \cite{opodis17-dvir} and Lee's HL1 \cite{Lee-HL1-TwoFace,icdcs05-lee}, 
when simplified for use in cache-coherent environments, both have extremely simple paths,
suggesting they wouild be competitive with Hemlock, but they 
do not readily tolerate multiple locks being held simultaneously.  
A crucial requirement for our design is that the lock algorithms can be used under
existing APIs such as pthread mutex locks or linux kernel locks, which allow multiple
locks to be held simultaneously and released in arbitrary order. 

\Invisible{Hemlock uses \emph{address-based} transfer of ownership, writing the address
of the lock instead of a boolean, making it different than HL1, MCS, etc.}

\section{Empirical Results}

Unless otherwise noted, all data was collected on an Oracle X5-2 system.
The system has 2 sockets, each populated with
an Intel Xeon E5-2699 v3 CPU running at 2.30GHz.  Each socket has 18 cores, and each core is 2-way
hyperthreaded, yielding 72 logical CPUs in total.  The system was running Ubuntu 20.04 with a stock
Linux version 5.4 kernel, and all software was compiled using the provided GCC version 9.3 toolchain
at optimization level ``-O3''.
64-bit C or C++ code was used for all experiments.
Factory-provided system defaults were used in all cases, and Turbo mode \cite{turbo} was left enabled.
In all cases default free-range unbound threads were used (no pinning of threads to processors). 

We implemented all user-mode locks within LD\_PRELOAD interposition
libraries that expose the standard POSIX \texttt{pthread\_\allowbreak{}mutex\_t} programming interface
using the framework from ~\cite{topc15-dice}.
This allows us to change lock implementations by varying the LD\_PRELOAD environment variable and
without modifying the application code that uses locks.
The C++ \texttt{std::mutex} construct maps directly to \texttt{pthread\_mutex} primitives,
so interposition works for both C and C++ code.
All lock busy-wait loops used the Intel \texttt{PAUSE} instruction.

\subsection{MutexBench benchmark} 
\label{Section:MutexBench} 



\begin{figure*}[ht]
\begin{minipage}[b]{0.45\linewidth}
\begin{adjustwidth}{-1cm}{}
\includegraphics[width=8.5cm]{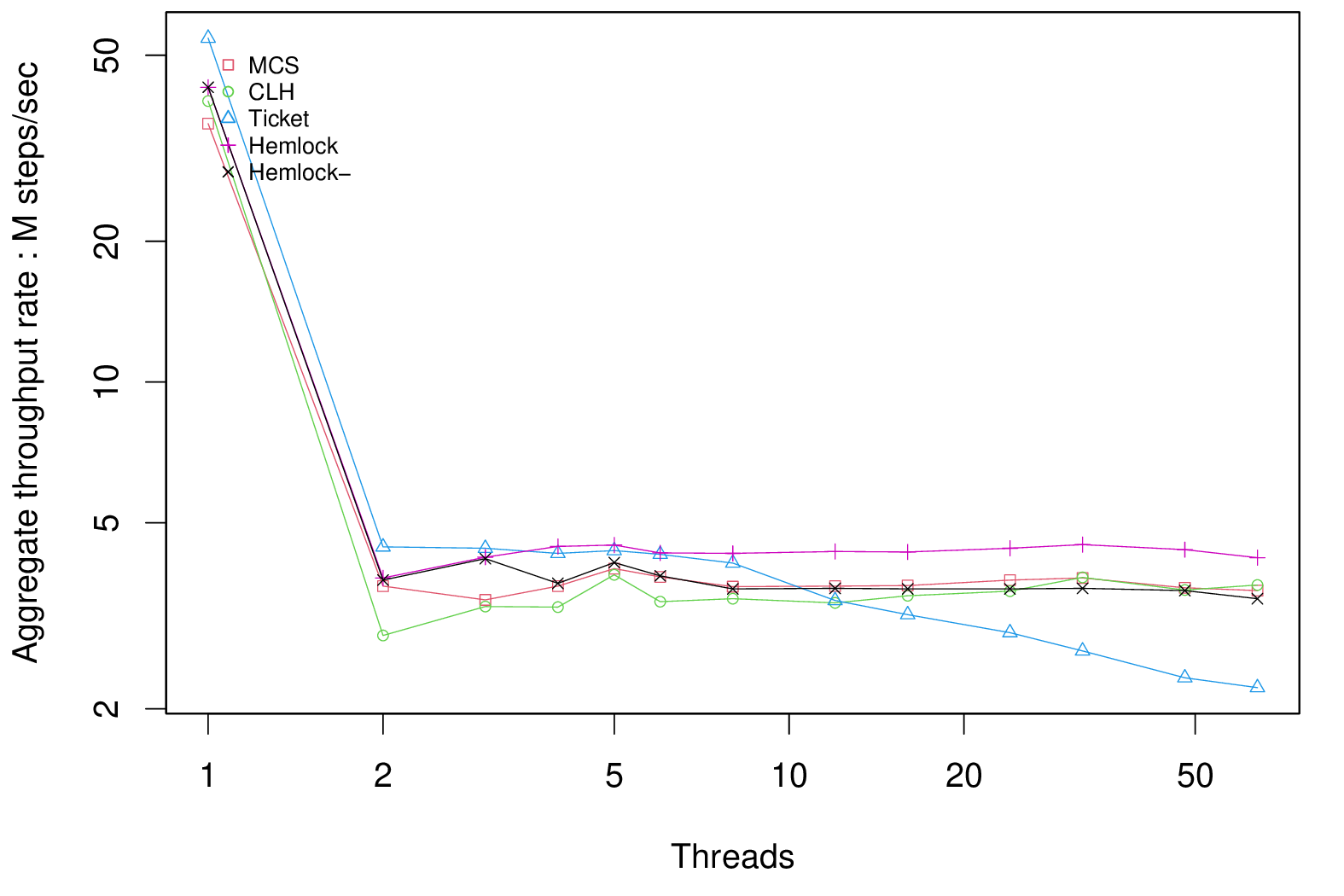}
\caption{MutexBench : Maximum Contention}
\label{Figure:MaximumContention}
\end{adjustwidth} 
\end{minipage}
\hspace{0.2cm}
\begin{minipage}[b]{0.45\linewidth}
\includegraphics[width=8.5cm]{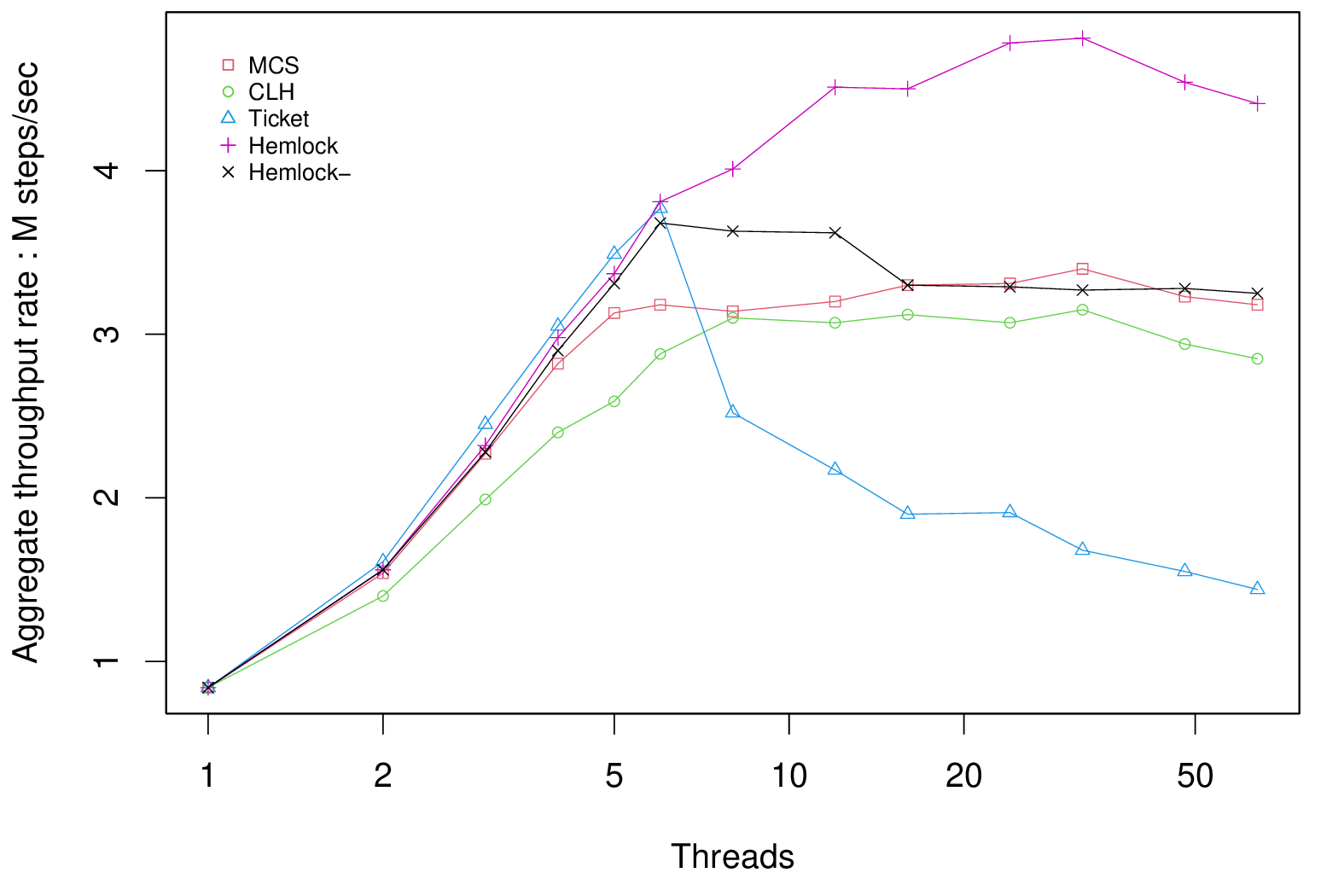}
\caption{MutexBench : Moderate Contention}
\label{Figure:ModerateContention}
\end{minipage}
\end{figure*}

\begin{figure*}[ht]
\begin{minipage}[b]{0.45\linewidth}
\begin{adjustwidth}{-1cm}{}
\includegraphics[width=8.5cm]{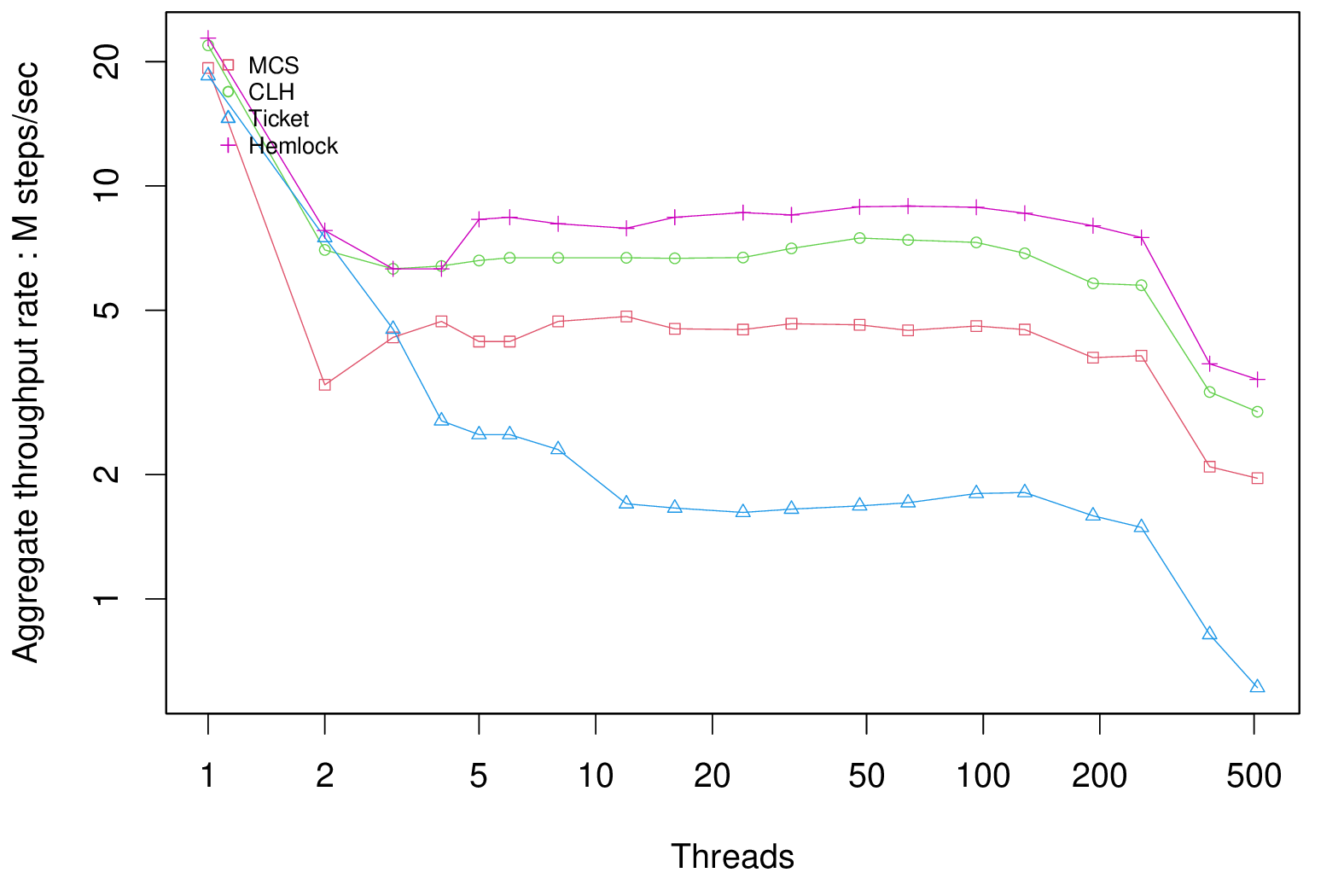}
\caption{MutexBench : Maximum Contention -- SPARC}
\label{Figure:MaximumContentionSPARC}
\end{adjustwidth} 
\end{minipage}
\hspace{0.2cm}
\begin{minipage}[b]{0.45\linewidth}
\includegraphics[width=8.5cm]{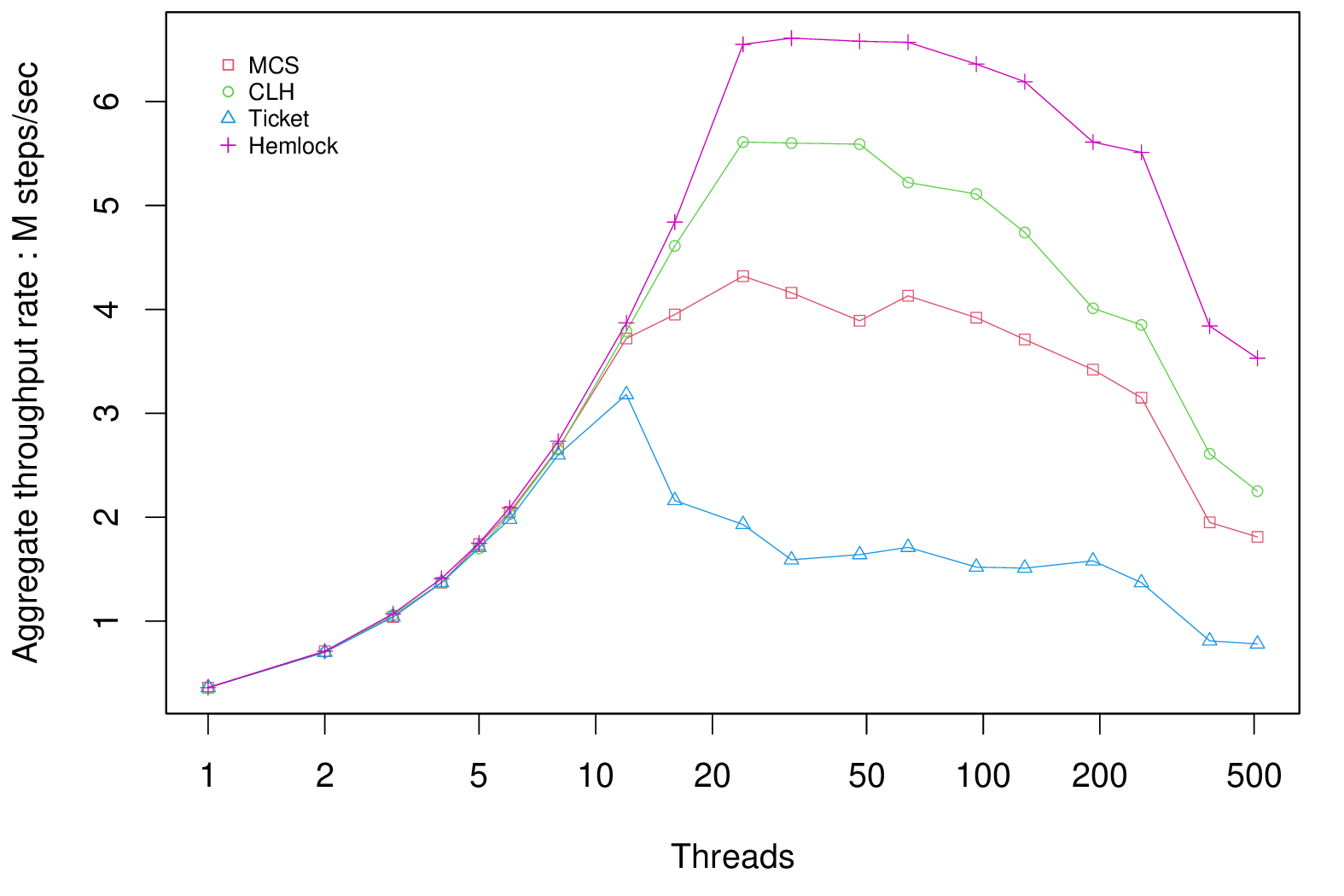}
\caption{MutexBench : Moderate Contention -- SPARC}
\label{Figure:ModerateContentionSPARC}
\end{minipage}
\end{figure*}

\begin{figure*}[ht]
\begin{minipage}[b]{0.45\linewidth}
\begin{adjustwidth}{-1cm}{}
\includegraphics[width=8.5cm]{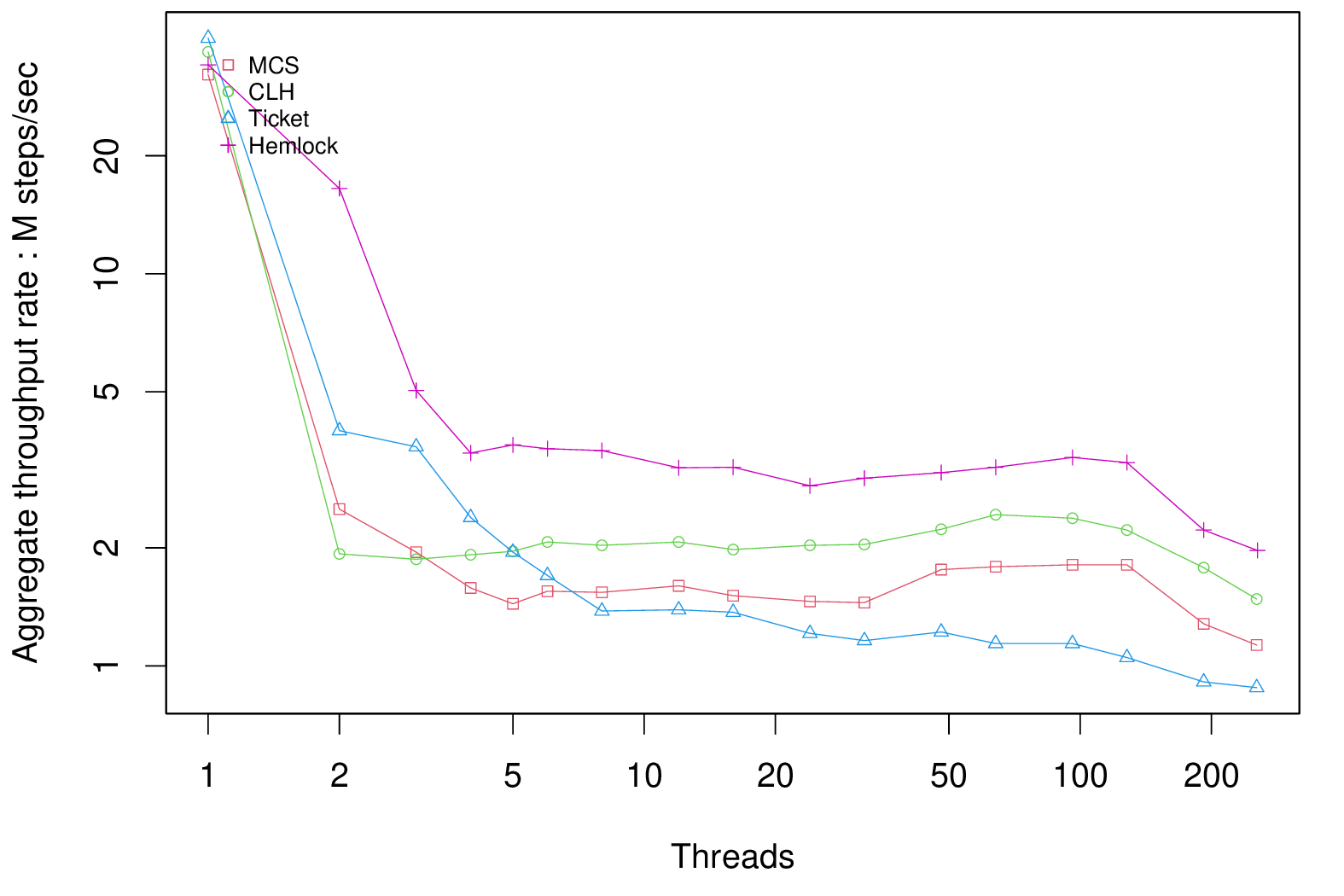}
\caption{MutexBench : Maximum Contention -- AMD}
\label{Figure:MaximumContentionAMD}
\end{adjustwidth} 
\end{minipage}
\hspace{0.2cm}
\begin{minipage}[b]{0.45\linewidth}
\includegraphics[width=8.5cm]{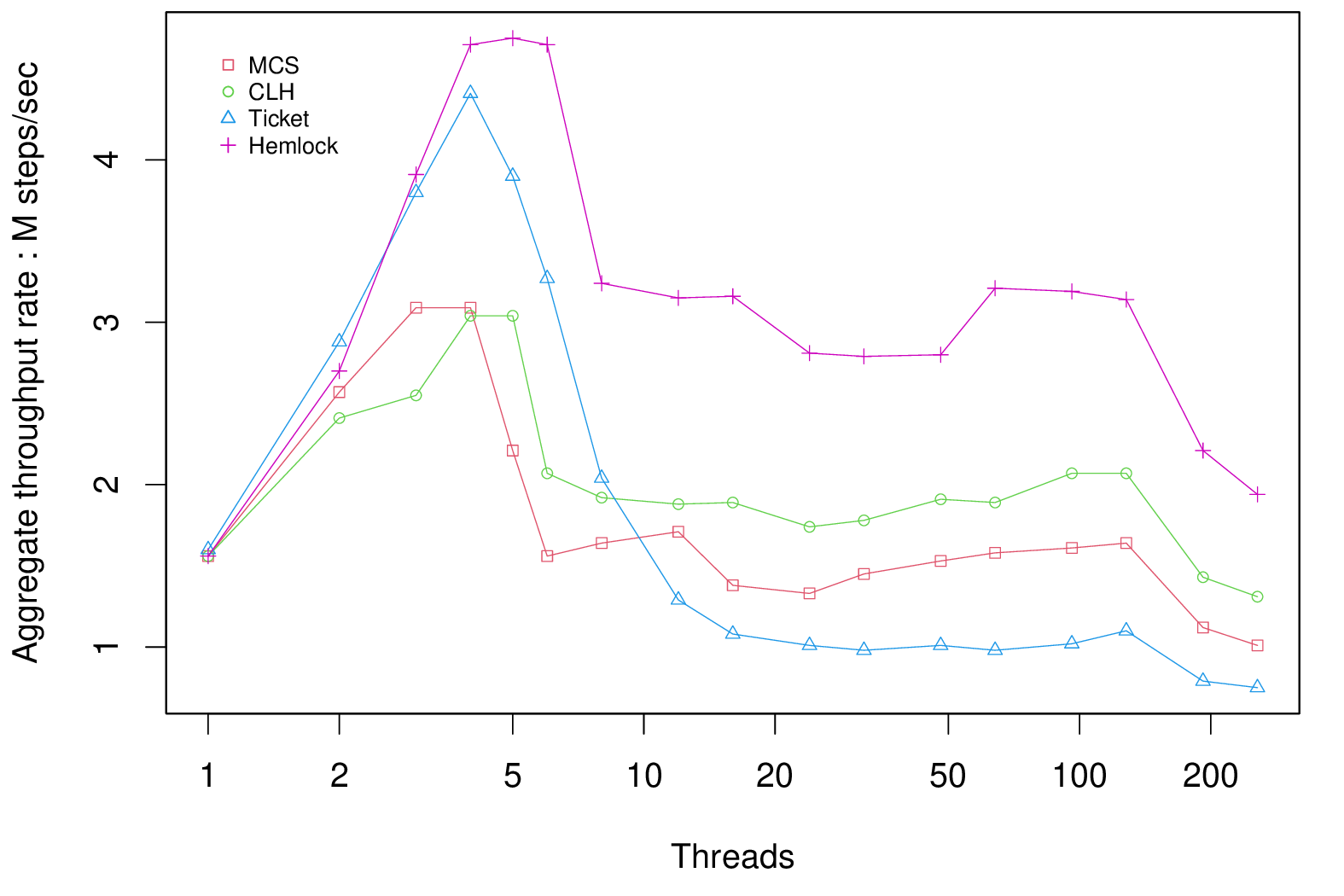}
\caption{MutexBench : Moderate Contention -- AMD}
\label{Figure:ModerateContentionAMD}
\end{minipage}
\end{figure*}

The MutexBench benchmark spawns $T$ concurrent threads. Each thread loops as follows:
acquire a central lock L; execute a critical section; release L; execute
a non-critical section. At the end of a 10 second measurement interval the benchmark
reports the total number of aggregate iterations completed by all the threads.
We report the median of 7 independent runs in Figure-\ref{Figure:MaximumContention}
where the critical section is empty as well as the non-critical section, subjecting the
lock to extreme contention.  (At just one thread, this configuration also constitutes
a useful benchmark for uncontended latency).  
The $X$-axis reflects the number of concurrently executing threads contending for the
lock, and the $Y$ reports aggregate throughput.
\Invisible{the tally of all loops executed by all the threads in the measurement interval} 
For clarity and to convey the maximum amount of information to allow a comparison of the algorithms,
the $X$-axis is offset to the minimum score and the $Y$-axis is logarithmic.

We ran the benchmark under the following FIFO/FCFS lock algorithms: 
\textbf{MCS} is classic MCS; 
\textbf{CLH} is CLH based on Scott's \emph{CLH variant with a standard interface} Figure-4.14 of \cite{Scott2013}; 
\textbf{Ticket} is a classic Ticket Lock;
\textbf{Hemlock} is the Hemlock algorithm, with the CTR optimization, described above.
\textbf{Hemlock-} is the naive Hemlock algorithm without the CTR optimization, and
correponds to Listing \ref{Listing:hemlock0}.   
For the MCS and CLH locks, our implementation stores the current head of 
the queue -- the owner -- in a field adjacent to the tail, so the lock body 
size was 2 words.  The Ticket Lock 
also has a size of 2 words, while Hemlock requires a lock body of just 1 word.  
MCS and CLH additionally require one queue element for each lock 
held or waited upon.  CLH also requires that each lock be initialized with a 
so-called dummy element.  To avoid memory allocation during the measurement interval,
the MCS implementation uses a thread-local stack of free queue elements 
\footnote{ As we are implementing a general purpose pthreads locking interface, a 
thread can hold multiple locks at one time.  Using MCS as an example, lets say thread 
$T1$ currently holds locks $L1$,$L2$,$L3$ and $L4$.  We'll assume no contention.
$T1$ will have deposited MCS queue nodes into each of those locks.  MCS nodes 
can not be reclaimed until the corresponding unlock operation.  Our implementaton 
could \texttt{malloc} and \texttt{free} nodes as necessary -- allocating in the 
lock operator and freeing in unlock -- but to avoid malloc and its locks, we 
instead use a thread-local stack of free queue nodes.  In the lock operator, 
we first try to allocate from that free list, and then fall back to \texttt{malloc}
only as necessary.  In unlock, we return nodes to that free list.  This approach 
reduces \texttt{malloc-free} traffic and the incumbent scalability concerns.   
We currently don't bother to trim the thread-local stack of free elements.  
So, if thread $T1$ currently holds no locks, the free stack will contain $N$ elements 
where $N$ is the maximum number of locks concurrently held by $T1$.  We 
reclaim the elements from the stack when $T1$ exits.  A stack is convenient for locality.}.  

In Figure-\ref{Figure:MaximumContention} we make the following observations regarding
operation at maximal contention with an empty critical section:
\footnote{We note in passing that care must be taken when \emph{negative} or \emph{retrograde}
scaling occurs and aggregate performance degrades as we increase threads.
As a thought experiment, if a hypothetical lock implementation were to introduce
additional synthetic delays outside the critical path, aggregate performance might increase as
the delay throttles the arrival rate and concurrency over the contended lock \cite{SIF}. 
As such, evaluating just the maximal contention case in isolation is insufficient.}.
\begin{itemize}[align=left,leftmargin=2em,labelwidth=0.8em]
\item At 1 thread the benchmark measures the latency of uncontended acquire and release operations.
Ticket Locks are the fastest, followed by Hemlock, CLH and MCS. 
\item As we increase the number of threads, Ticket Locks initially do well but then fade,
exhibiting a precipitous drop in performance.
\item Broadly, Hemlock performs slightly better than or the same as CLH or MCS. 
\end{itemize} 

To gauge the contribution and benefit of the CTR optimization, we can compare \texttt{Hemlock},
which incorporates CTR, against
\texttt{Hemlock-}, the simplistic reference implementation, shown in Listing-\ref{Listing:hemlock0}.


\Invisible{
Sensitivity analysis : breakdown and contribution of AH and CTR optimizations. 
Threat-to-validity and confounding factor : 
atomics in CTR might result in stall or delay and result in SIF effect : slower-is-faster.
Cross-check with variants that use non-CTR spinning but with dummy atomics.  
}

In Figure-\ref{Figure:ModerateContention} we configure the benchmark so the non-critical section
generates a uniformly distributed random value in $[0-400)$ and steps a
thread-local C++ \texttt{std::mt19937} random number generator (PRNG) that many steps, 
admitting potential positive scalability.  The critical section advances a shared random 
number generator 5 steps. 
In this moderate contention case we observe that Ticket Locks again do well at low thread counts,
and that Hemlock outperforms both MCS and CLH.

\Invisible{Maximum extreme contention -- maximizes arrival rate} 

\subsection{MutexBench Benchmark : SPARC} 

To show that our approach is general and portable, we next report MutexBench results on a Sun/Oracle T7-2
\cite{sparc-T7-2} in Figures \ref{Figure:MaximumContentionSPARC} and \ref{Figure:ModerateContentionSPARC}.  
The T7-2 has 2 sockets, each socket populated by an M7 SPARC CPU running at 4.13GHz
with 32 cores.  Each core has 8 logical CPUs sharing 2 pipelines.  The system has 512 logical CPUs and 
was running Solaris 11.  
We used the GCC version 6.1 toolchain to compile the benchmark and the lock libraries.  
64-bit SPARC does not directly support atomic \texttt{fetch-and-add} or \texttt{SWAP} 
operations -- these are emulated by means of a 64-bit compare-and-swap operator (\texttt{CASX}).  
To implement CTR in the waiting phase, we used \texttt{MONITOR-MWAIT} on the predecessor's 
\texttt{Grant} field followed by an immediate \texttt{CASX} to try to reset \texttt{Grant},  
avoiding the promotion from \emph{shared} to \emph{modified} state which would normally be
found in naive busy-waiting.  
As needed, \texttt{CASX(A,0,0)} serves as the read-with-intent-to-write primitive. 
The system uses MOESI cache coherency instead of the MESIF \cite{MESIF} found in modern 
Intel-branded processors, allowing more graceful handling of write sharing.  
The abrupt performance drop experienced by all locks
starting at 256 threads is caused by competition for pipeline resources.  

\subsection{MutexBench Benchmark : AMD} 
Figures \ref{Figure:MaximumContentionAMD} and \ref{Figure:ModerateContentionAMD}
show performance on a 2-socket AMD NUMA system, where each socket contains an 
EPYC 7662 64-Core Processor and each core supports 2 logical CPUs, for 256 logical processors
in total.  The base clock speed is 2.0 GHz.  The kernel was linux version 5.4 and we used the same 
binaries built on the Intel X5-2 system.  
AMD uses a MOESI coherence protocol.
The results on AMD concur with those observed on the Intel system. 


\subsection{LevelDB} 

\begin{figure}[h]                                                                     
\includegraphics[width=8.5cm]{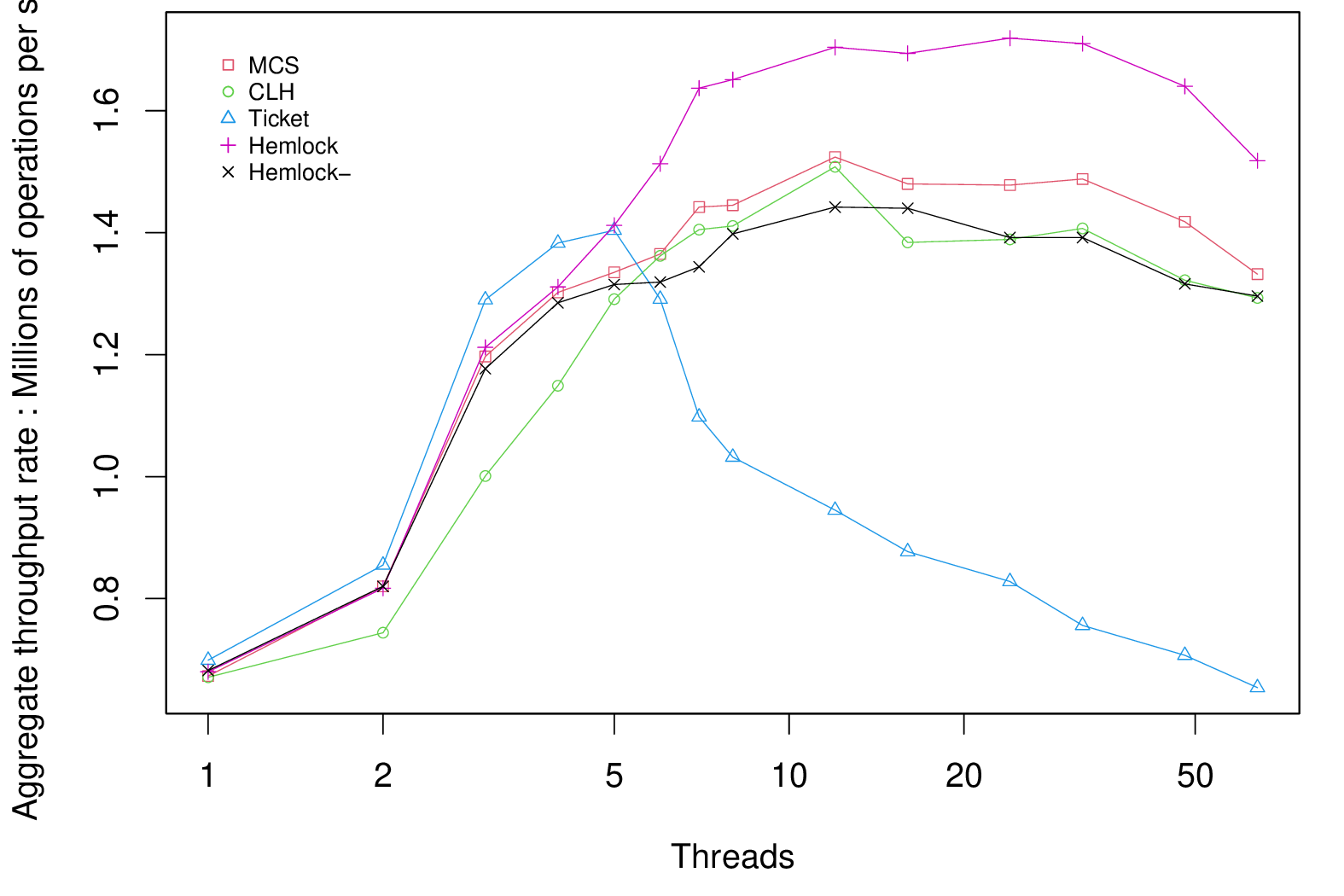}
\caption{LevelDB readrandom}                                                   
\label{Figure:readrandom}                                                                  
\end{figure}                     


In Figure-\ref{Figure:readrandom}  we used the ``readrandom'' benchmark in LevelDB version 1.20 
database \footnote{\url{leveldb.org}} varying the number of threads and reporting throughput 
from the median of 5 runs of 50 second each.  
Each thread loops, generating random keys and then tries to read the associated value from
the database.  
We used the Oracle X5-2 system to collect data.  
We first populated a database \footnote{db\_bench \---\---threads=1 
\textendash{}\textendash{}benchmarks=fillseq \---\---db=/tmp/db/}
and then collected data \footnote{db\_bench \---\---threads=\emph{threads}  
\---\---benchmarks=readrandom \\ \---\---use\_existing\_db=1 
\---\---db=/tmp/db/ \---\---duration=50}. 
We made a slight modification to the \texttt{db\_bench} benchmarking 
harness to allow runs with a fixed duration that reported aggregate throughput.  
Ticket Locks exhibit a slight advantage over MCS, CLH and Hemlock at low threads count after
which Ticket Locks fade. 
LevelDB uses coarse-grained locking, protecting the database with a single central mutex: 
\texttt{DBImpl::Mutex}.  Profiling indicates contention on that lock via \texttt{leveldb::DBImpl::Get()}.  

Using an instrumented version of Hemlock we characterized the application behavior
of LevelDB, as it relates to Hemlock.  At 64 threads, during a 50 second run,
we found 24 instances of 
calls to \texttt{lock} where a thread already held at least one other lock. 
These all occurred during the first second after startup.  
The maximum number of locks held simultaneously by any thread was 2.    
The maximum number of threads waiting simultaneously on any \texttt{Grant} field was 1, 
thus the application enjoyed purely local spinning.  


\subsection{Impact of CTR Optimization} 
\label{Section:OffCore} 


\Invisible{
CTR benefit : Sensitivity analysis -- show correlation between offcore traffic and performance.
Support our claims about efficacy of CTR and mode-of-benefit
}

We used the built-in linux \texttt{perf stat} command to collect data from the hardware performance 
monitoring unit counters and found that CTR reduced total \texttt{offcore} traffic \cite{IntelUncore},
while providing an improvement in throughput.
Table \ref{Table:OffCore} examines the execution of the \texttt{MutexBench} benchmark on the X5-2
system configured for 32 threads and with 0-length critical and non-critical sections. 
The \emph{Rate} column is given in units of millions of lock-unlock operations completed per second
and the \emph{OffCore} column reports the number of offcore accesses
\footnote{we used the sum of \texttt{offcore\_requests.all\_data\_rd} and 
\texttt{offcore\_requests.demand\_rfo}} counters per lock-unlock pair.  Offcore accesses 
are memory references that can not be satisifed from the core's local L2 cache, including
coherence misses.  As the working set of the each thread in the benchmark is tiny, offcore
accesses largely reflect cache coherent communications arising from acquiring and releasing
the lock.  As we can see Hemlock with CTR yields higher throughput than Hemlock without CTR,
and incurs less offcore traffic.  Both CLH and MCS suffer from moderately elevated
offcore communication rates.  We isolated that increase to the stores the reinitialize
the queue nodes in preparation for reuse.  Those stores execute outside the critical section.


\begin{table} [h]
\centering
\begin{tabular}{lcc}
\toprule
\multicolumn{1}{l}{Lock} &
\multicolumn{1}{c}{Rate} &
\multicolumn{1}{c}{OffCore}\\
\midrule

MCS                  & 3.81   & 10.6  \\
CLH                  & 3.82   & 11.1  \\
Ticket Locks         & 2.66   & 45.9  \\
Hemlock              & 4.48   & 6.81  \\
Hemlock without CTR  & 3.62   & 7.92  \\ 

\midrule[\heavyrulewidth]
\bottomrule
\end{tabular}%
\caption{Impact of CTR on OffCore Access Rates}\label{Table:OffCore}
\end{table}

\Invisible{We also found that CTR resulted in a reduction in number of load operations
that ``hit'' on a line in $M$-state in another core's cache -- requiring write invalidation
and transfer back to the requester's cache.} 

We can show similar benefits from CTR with a simple program where a set of concurrent threads are 
configured in a ring, and circulate a single token.  A thread waits for its mailbox to become
non-zero, clears the mailbox, and deposits the token in its successor's mailbox.  
Using \texttt{CAS}, \texttt{SWAP} or \texttt{Fetch-and-Add} to busy-wait improves the circulation 
rate as compared to the naive form which uses loads.

\subsection{Multi-waiting} 

\begin{figure}[h]                                                                     
\includegraphics[width=8.5cm]{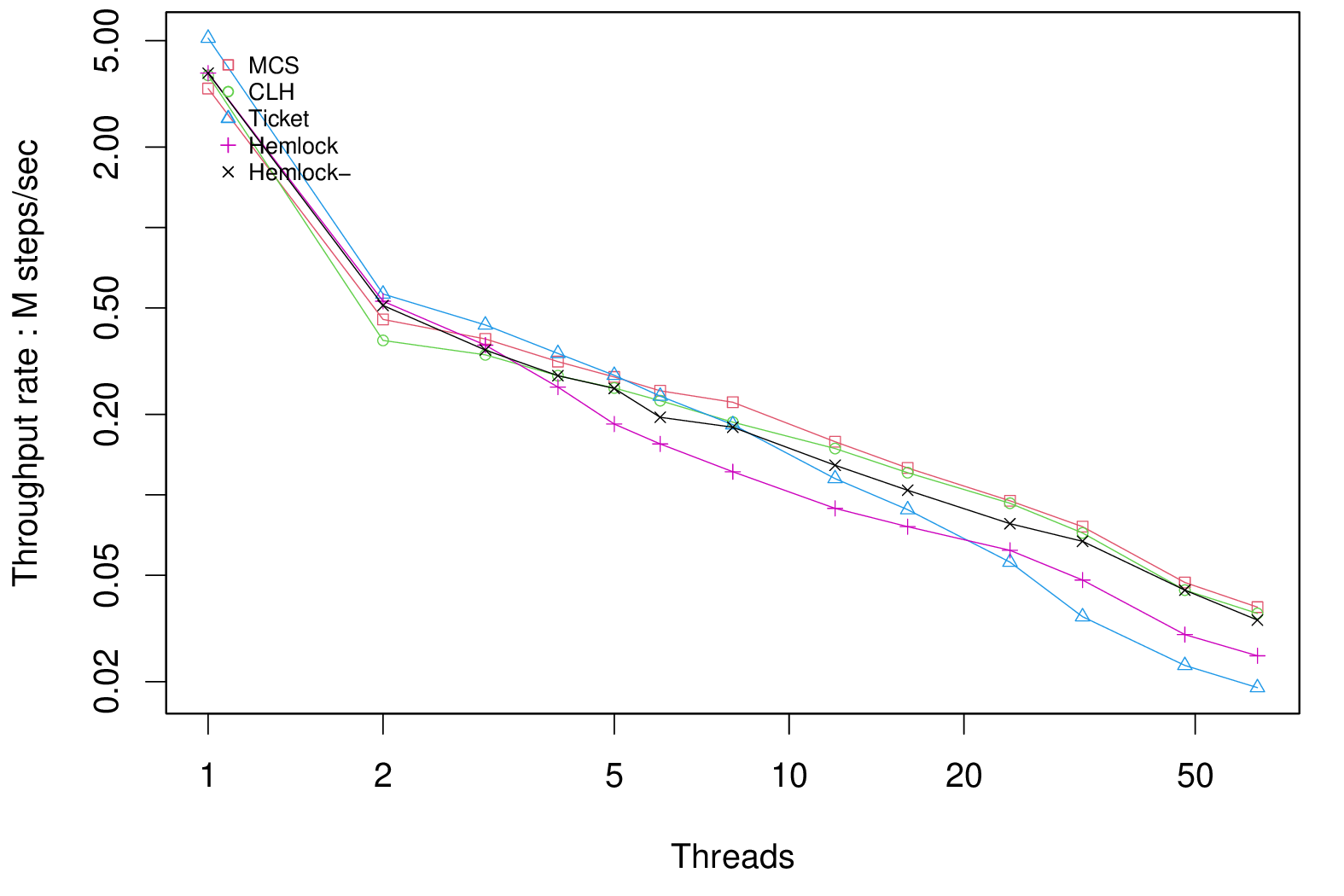}
\caption{Multi-waiting}                                                   
\label{Figure:multiwait}                                                                  
\end{figure}

\Invisible{Constructed; devised; contrived; conjured} 

We intentionally constructed a benchmark that induces multi-waiting to measure the
performance of Hemlock in a challenging and unfavorable operating region.  
We modify \texttt{MutexBench} to have an array of 10 shared locks. There is a single dedicated
``leader'' thread which loops as follows : acquire all 10 lock in ascending order and then 
release the locks in reverse order.  At the end of the measurement interval the leader
reports the number of steps it completed, where a step consists of acquiring and releasing all the locks   
All the other threads loop, picking a single random lock from the set of 10, and 
then acquire and release that lock.   We ignore the number of iterations completed by 
the non-leader threads.   Neither the leader nor the non-leaders execute any delays in their 
critical or non-critical phases.  When configured for 32 threads, for example, we have
1 leader and 31 non-leaders.  
In general, the worst-case maximum number of threads busy-waiting on a given location at a given time
is as follows : $1$ for CLH and MCS, which enjoy purely local spinning; 
$T-1$ for ticket locks; and $min(T-1,N-1)$ for Hemlock, where $T$ is the number 
of threads and $N$ is the number of locks, which is 10 in our configuration.  
We plot the throughput results in Figure-\ref{Figure:multiwait}.  Data for this
experiment was collected on the Oracle X5-2 described earlier.  

As we increase the number of threads, performance, as expected, drops over all the lock
algorithms as the primary leader threads suffers more obstruction from the non-leader threads.  
And as usual, ticket lock performs well, relative to other locks, at low thread counts 
but the performance then falls behind as we increase the number of threads. 
Hemlock-, without CTR, performs somewhat
worse than CLH and MCS as we increase the number of threads, and multi-waiting
increases.  Finally, Hemlock with CTR performs worse than Hemlock- as the CTR form
optimistically assumes multi-waiting is rare and busy-waits in an impolite fashion 
with \texttt{CAS} instead of loads.  As such, a \texttt{grant} field subject to multi-waiting
will \emph{slosh} or bounce between caches as each waiting thread drives the underlying
line into exclusive $M$-state.  This behavior consumes interconnect bandwidth and can 
retard lock ownership handover.  The CTR optimization is actually harmful under
high degrees of multi-waiting.

\section{Future Work} 


An interesting variation we intend to explore in the future is to replace the 
simplistic spinning on the \texttt{Grant} field with a per-thread condition variable and mutex pair that 
protect the Grant field, allowing threads to use the same waiting policy as the platform
mutex and condition variable primitives.  All long-term waiting for the \texttt{Grant} field
to become a certain address or to return to $0$ would be via the condition variable.
Essentially, we treat \texttt{Grant} as a bounded buffer of capacity 1 protected
in the usual fashion by a condition variable and mutex.
This construction yields 2 interesting properties : (a) the new lock 
enjoys a fast-path, for uncontended locking, that doesn’t require any underlying mutex 
or condition variable operations, 
(b) even if the underlying system mutex isn’t FIFO, our new lock provides strict FIFO admission. 
Again, the result is compact, requiring only a mutex, condition variable and \texttt{Grant} field
per thread, and only one word per lock to hold the \texttt{Tail}.  For systems where 
locks outnumber threads, such an approach would result in space savings.

\section{Conclusion} 

Hemlock trades off improved space complexity against the cost of higher remote memory 
reference (RMR) complexity. 
Hemlock is exceptionally simple with short paths, and avoids the dependent loads and
indirection required by CLH or MCS to locate queue nodes.  
The contended handover critical path is extremely short --
the unlock operator conveys ownership to the successor in an expedited fashion.  
Despite being compact, it provides 
local spinning in common circumstances and scales better than Ticket Locks.  
Instead of traditional queue elements, as found in CLH and MCS, we use
a per-thread shared singleton element.  
Finally, Hemlock is practical and readily usable in real-world lock implementations. 

\begin{acks}                          
We thank Peter Buhr and Trevor Brown at the University of Waterloo for access to their AMD system.  
\end{acks}

\bibliography{hemlock}

\newpage        

\appendix


\section{Optimization: Overlap}

To reduce the impact of waiting for receipt of transfer in the unlock operator, 
at Listing-\ref{Listing:hemlock0} line 20,
we can apply the \emph{Overlap} optimization which shifts and defers that waiting step until 
subsequent synchronization operations, allowing greater overlap between the successor and
the outgoing owner. 

Threads arriving in the lock operator at Listing-\ref{Listing:hemlockOVERLAP} line 6 wait to ensure 
their \texttt{Grant} mailbox field does not contain a residual address from a previous contended
unlock operation on that same lock, in which case it must wait for that tardy
successor to fetch and clear the \texttt{Grant} field
\footnote{We thank Adrian Uffmann for identifying an error in earlier versions of this figure.}. 
In practice, waiting on this condition is rare. 
(If thread $T1$ were to enqueue an element that contains an residual \texttt{Grant} value that happens to 
match that of the lock, then when a successor $T2$ enqueues after $T1$, it will incorrectly see 
that address in T1’s grant field and then incorrectly enter the critical section, resulting in 
exclusion and safety failure and a corrupt chain.  The check at line 6 prevents that pathology).  

In Listing-\ref{Listing:hemlockOVERLAP} line 16, threads wait for their own \texttt{Grant} field to become empty.  
\texttt{Grant} could be non-\texttt{null} because of previous unlock operations that wrote
an address into the field, but the corresponding successor has not yet cleared the field back 
to \texttt{null}.  That is, \texttt{Grant} is still occupied.   
Once \texttt{Grant} becomes empty, the thread then writes the address of the
lock into \texttt{Grant}, alerting the successor and passing ownership.
When ultimately destroying a thread, it is then necessary to wait for the thread's \texttt{Grant} field
to transition back to \texttt{null}  before reclaiming the memory underlying \texttt{Grant}.  
This can be accomplished by waiting in the thread's destructor method.  

\lstset{language=Python}
\lstset{frame=lines}
\lstset{caption={Hemlock with Overlap Optimization}}
\lstset{label={Listing:hemlockOVERLAP}}
\lstset{basicstyle=\footnotesize\ttfamily} 
\lstset{commentstyle=\itshape\color{gray}} 
\lstset{commentstyle=\slshape\color{gray}} 
\lstset{commentstyle=\itshape\color{gray}} 
\lstset{keywordstyle=\color{forestgreen}\bfseries} 
\lstset{backgroundcolor=\color{Gray95}} 
\begin{listing}[htp]    
\begin{adjustwidth}{-0em}{0pt}
\begin{lstlisting}[language=Python,mathescape=true,escapechar=\%]
class Thread :
  atomic<Lock *> Grant = null 

class Lock :
  atomic<Thread *> Tail = null 

def %\underline{Lock}% (Lock * L) :
  while Self$\rightarrow$Grant $==$ L : Pause
  auto pred = swap (&L$\rightarrow$Tail, Self) 
  if pred $\neq$ null :
     while pred$\rightarrow$Grant $\neq$ L : Pause
     pred$\rightarrow$Grant = null
  assert L$\rightarrow$Tail $\neq$ null
  
def %\underline{Unlock}% (Lock * L) :  
  auto v = cas (&L$\rightarrow$Tail, Self, null)
  assert v $\neq$ null
  if v $\neq$ Self :
     while Self$\rightarrow$Grant $\neq$ null : Pause
     Self$\rightarrow$Grant = L 

\end{lstlisting}
\end{adjustwidth}
\end{listing}

\section{Optimization: Aggressive Hand-Over}  
\label{AHO} 

\lstset{label={Listing:hemlockAH}}
\lstset{caption={Hemlock with Aggressive Hand-Over Optimization}}
\begin{listing}[htp]    
\begin{adjustwidth}{-0em}{0pt}
\begin{lstlisting}[language=Python,mathescape=true,escapechar=\%]
class Thread :
  atomic<Lock *> Grant = null 

class Lock :
  atomic<Thread *> Tail = null 

def %\underline{Lock}% (Lock * L) :
  assert Self$\rightarrow$Grant $==$ null
  auto pred = swap (&L$\rightarrow$Tail, Self) 
  if pred $\neq$ null :
     while cas(&pred$\rightarrow$Grant, L, null) $\neq$ L : Pause
  
def %\underline{Unlock}% (Lock * L) :  
  assert Self$\rightarrow$Grant $==$ null
  Self$\rightarrow$Grant = L 
  auto v = cas (&L$\rightarrow$Tail, Self, null)
  if v $=$ Self :
    Self$\rightarrow$Grant = null
    return
  while FetchAdd(&Self$\rightarrow$Grant, 0) $\neq$ null : Pause

\end{lstlisting}
\end{adjustwidth}
\end{listing}

The \emph{Aggressive Hand-Over (AH)} optimization, shown in Listing-\ref{Listing:hemlockAH}, 
changes the code in \texttt{unlock} to first store the lock's address into the \texttt{Grant} field
(Listing-\ref{Listing:hemlockAH} Line 12), optimistically anticipating the existence of waiters, 
and then execute the atomic \texttt{CAS} to try to swing the \texttt{Tail} field back
from \texttt{Self} to \texttt{null}, handling the uncontended case. 
If the \texttt{CAS} succeeded, there are no waiters, and we then  
reset \texttt{Grant} back to \texttt{null} and return, 
and otherwise wait for the successor to clear \texttt{Grant}. 
This reorganization accomplishes handover earlier in the \texttt{unlock} path and improves scalability
by reducing the critical path for handover.  
Handover time impacts the scalability as the lock is held throughout handover, increasing the effective
length of the critical section \cite{isca10-eyerman,podc18-aksenov}. 
For uncontended locking, where there are no waiting successors, the superfluous stores to set 
and clear \texttt{Grant} are harmless to latency as the thread is likely to have the 
underlying cache line in \emph{modified} state in its local cache. 
Listing-\ref{Listing:hemlockAH} also incorporates the CTR optimization.  

The contended handover critical path is extremely short --
the very first statement in the unlock operator, at line 12, conveys ownership to the successor. 

In \texttt{unlock}, after we store into the \texttt{Grant} field and transfer
ownership, the successor may enter the critical section and even release the lock 
in the interval before the original owner reaches the \texttt{CAS} in \texttt{unlock}. 
As such, it is possible that the \texttt{CAS} in \texttt{unlock} could fetch 
a \texttt{Tail} value of \texttt{null}.  We therefore remove the corresponding \code{assert} 
found in line 17 in Listing-\ref{Listing:hemlock0}.  

\Invisible{In a sense, when we have waiters and contention, executing the \texttt{CAS} first in 
\texttt{unlock} adds useless latency and coherence traffic, and delays the 
handover to the successor.
} 

\Invisible{Optimistic; Opportunistic; Early; Eager; Agro; Anticipatory; 
speculative; Accelerate; Preemptive} 

While the aggressive hand-over optimization improves contended throughput, 
it can lead to surprising \emph{use-after-free} memory lifecycle pathologies and
is thus not safe for general use in a \texttt{pthread\_mutex} implementation  
\footnote{We thank Alexander Monakov and Travis Downs for reminding us of this concern}. 

Consider the following scenario where we have a structure instance $I$ that contains a
lock $L$ and a reference count for $I$. 
The reference count, which is currently $2$, is protected by $L$. 
Thread $T1$ currently holds $L$ while it accesses $I$.  
Thread $T2$ arrives and stalls waiting to acquire $L$ and access $I$.  
$T1$ finishes accessing $I$, decrements the reference count from $2$ to $1$ and then 
calls \texttt{unlock(L)}.  $T1$ executes Listing-\ref{Listing:hemlockAH} line 12 and
then stalls.  $T2$ then acquires $L$ and accesses $I$.  
When finished, $T2$ reduces the reference count from $1$ to $0$, making note of that fact.
$T2$ then releases $L$, and, as the reference count transitioned to $0$, and $I$ should
not longer be accessible or reachable, $T2$ frees the memory associated with $I$, which
includes $L$.  $T1$ -- a tardy straggler -- resumes at line 13 and accesses $L$, resulting in a use-after-free error.
Similar pathologies have been observed and fixed in the linux kernel lock implementation and the
user-mode \texttt{pthread\_mutex} implementations \cite{LifecycleSurprise1,LifecycleSurprise2}. 

Broadly, if the unlock operator has a fast-path which might release or transfer the lock, and, 
in the same invocation of unlock, might then subsequently access the lock body, then the lock 
implementation is exposed to the use-after-free problem.  
Put another way, once transfer has been effected or potentially effected, the \texttt{unlock} implementation
must not access the lock body again.  
In our case, the speculative hand-over store at line 12 renders the AH algorithm vulnerable. 

AH remains safe and immune from use-after-free errors, however, in any environment where the 
lock body $L$ can not recycle while a thread remains in \texttt{unlock(L)}.  
In garbage-collected environments, or where $L$ is protected by safe memory reclamation techniques (SMR) -- such as 
\emph{read-copy update} (RCU), Hazard Pointers, or Epoch-based Reclamation -- then AH is 
permissible as the thread calling \texttt{Unlock(L)} 
continues to hold a reference to $L$ which prevents $L$ from recycling.  
Furthermore, AH is also safe if $L$ resides in type-stable memory or if $L$ is never deallocated,
as would be the case for statically allocated locks.  

The AH form (with CTR) provides the best overall performance of the Hemlock family
and is our preferred form when lifecycle concerns permit.  

As a general rule, if there are no waiters, no possiblity of new arrivals, and the lock is not held,
then it is safe to call the destructor and then subsequently recycle the memory underlying the lock. 
Absent AH, HemLock lock instances are \emph{trivially destructible}
\footnote{\url{https://en.cppreference.com/cpp/language/destructor\#Trivial\_destructor}}
and we say the lock is \emph{prompt destruction safe}.  
With AH, however, for general usage, additional safety precautions are necessary for safe destruction,
as the lock does not provide prompt destruction safety, and the is not trivially destructible.  
As noted above, SMR can be used.  

Another approach to allow AH is to augment the lock with an 
atomic reference counter, which is incremented on arrival in \texttt{lock} and decremented as the last step
as threads depart \texttt{unlock}.  In turn, the destructor will wait for the count to drop
to zero before returning, delaying until tardy threads fully depart \texttt{unlock}, 
and avoiding use-after-free errors. 
We also observe that such an atomic reference counter itself causes a coherence hotspot and
impedes the scalability of a lock so augmented. 
Furthermore, some locking APIs do not even expose destructors, such
as the linux kernel \texttt{qspinlock} construct, 
and that applications do not bother to call destructors even if they exist, as they 
assume the lock implementation will be prompt destruction safe. 
Using gcc++-15 on ubuntu 26.04 x86, where \texttt{std::mutex} is implemented via the 
POSIX \texttt{pthread\_mutex} subsystem,  we find the compiler elides calls to \texttt{pthread\_mutex\_destroy}
for \texttt{std::mutex} instances  
\footnote{\texttt{std::is\_trivially\_destructible\_v<std::mutex>} reports \texttt{true}}.  
We believe this reflects an unwarranted assumption in the C++ \texttt{std::mutex} libraries which
incorporates and embodies knowledge of the current specific linux \texttt{pthread\_mutex} implementation. 
This behavior can be overriden by defining the pre-processor
symbol \texttt{\_GTHREAD\_USE\_MUTEX\_INIT\_FUNC} which causes the underlying \texttt{pthread\_mutex} 
constructors and destructors
to be invoked when \texttt{std::mutex} instances are created and destroyed, 
but all relevant code within the process -- including C++ code in dynamic shared objects -- must be recompiled.
Eliding constructor and destructor calls in this fashion may improve performance, but precludes the safe use of 
\texttt{LD\_PRELOAD} interposition for lock algorithms that require non-trivial constructors or destructors.  
CLH, for instance, could not be safely implemented under \texttt{std::mutex} via \texttt{LD\_PRELOAD} 
interposition on the \texttt{pthread\_mutex} operators, as failure to call destructors
constitutes a memory leak of the queue element currently associated with the lock instance. 

\Invisible{Favorite; preferred embodiment} 

\Invisible{Even if a lock recycles, I think we’re safe under TSM as there’s no way the CAS 
(following the speculative store into L->Grant in unlock) could find the caller’s thread 
in the tail field and do any damage.}  

\lstset{label={Listing:hemlockOHOvA}}
\lstset{caption={Hemlock with Optimized Hand-Over -- Variant 1}}
\begin{listing}[htp]    
\begin{adjustwidth}{-0em}{0pt}
\begin{lstlisting}[language=Python,mathescape=true,escapechar=\%]
class Thread :
  atomic<Lock *> Grant = null 

class Lock :
  atomic<Thread *> Tail = null 

def %\underline{Lock}% (Lock * L) :
  auto pred = swap (&L$\rightarrow$Tail, Self) 
  if pred $\neq$ null :
     ## Note that we do use the value returned by the following cas() 
     cas (&pred$\rightarrow$Grant, 0, L|1) ; 
     while cas(&pred$\rightarrow$Grant, L, null) $\neq$ L : Pause
  
def %\underline{Unlock}% (Lock * L) :  
  if Self$\rightarrow$Grant $==$ (L|1) :
     PassLock: 
     Self$\rightarrow$Grant = L 
     while FetchAdd(&Self$\rightarrow$Grant, 0) $==$ L : Pause
     return
  auto v = cas (&L$\rightarrow$Tail, Self, null)
  assert v $\neq$ null 
  if v $\neq$ Self : goto PassLock

\end{lstlisting}
\end{adjustwidth}
\end{listing}

We now show additional variants that avoid use-after-free concerns, but
which still provide the fast contended hand-over exhibited by AH.

In Lisiting-\ref{Listing:hemlockOHOvA} 
we augment the encoding of \texttt{Grant} to
add a distinguished $L|1$ state, borrowing the low-order bit of the lock address (which is otherwise 0) 
as a flag to indicate that a successor exists.   In the unlock operator, if a thread discovers 
that its \texttt{Grant} field is $L|1$ then it is certain that an immediate successor
exists for $L$, in which case the thread overwrites $L|1$ with $L$ to pass ownership to
that sucessor.  This approach also avoids, for common modes of contention, any accesses
to the lock's \texttt{Tail} field in the \texttt{unlock} operator, further reducing coherence traffic
on that coherence hotspot. (Both MCS and CLH, under steady-state sustained contention, and 
assuming that context is passed by means other than fields in the lock body, typically manage to
avoid accessing the lock body in the \texttt{unlock} operator, a property which benefits perforamnce.) 
By eliminating the speculative store into \texttt{Grant} found
in AH, we avoid use-after-free concerns.

\lstset{label={Listing:hemlockOHOvB}}
\lstset{caption={Hemlock with Optimized Hand-Over -- Variant 2}}
\begin{listing}[htp]    
\begin{adjustwidth}{-0em}{0pt}
\begin{lstlisting}[language=Python,mathescape=true,escapechar=\%]
class Thread :
  atomic<Lock *> Grant = null 
  atomic<Lock *> Waiting = null 

class Lock :
  atomic<Thread *> Tail = null 

def %\underline{Lock}% (Lock * L) :
  Self$\rightarrow$Waiting $=$ null
  auto pred = swap (&L$\rightarrow$Tail, Self) 
  if pred $\neq$ null :
     pred$\rightarrow$Waiting $=$ L 
     while pred$\rightarrow$Grant $\neq$ L : Pause
     pred$\rightarrow$Grant $=$ null 
     
  
def %\underline{Unlock}% (Lock * L) :  
  ## We can optionally apply the AH optimization to 
  ## this variant, if designed, by agressively and 
  ## opportunistically storing L into Self->Grant 
  ## on arrival in unlock() and then
  ## clearing Self->Grant as needed in the
  ## uncontended path
  if Self$\rightarrow$Waiting $==$ L :
     ## We have an visible waiter and can effect
     ## efficient handover without accessing the L body. 
     ## Note that we do not need to explicitly clear 
     ## the Self->Waiting field, although doing so is 
     ## safe and provides better hygiene and allows 
     ## better assert()s and diagnostics
     PassLock: 
     Self$\rightarrow$Grant = L 
     while Self$\rightarrow$Grant $==$ L : Pause
     return
  ## apparently uncontended ... 
  auto v = cas (&L$\rightarrow$Tail, Self, null)
  assert v $\neq$ null 
  if v $\neq$ Self : goto PassLock

\end{lstlisting}
\end{adjustwidth}
\end{listing}

In Listing-\ref{Listing:hemlockOHOvB} we provide an alternative to Listing-\ref{Listing:hemlockOHOvA}.
This variant eliminates the special distinguished \emph{marked} $L|1$ encoding and also avoids the use of 
read-modify-write atomic operations, but at the expense of adding a new \texttt{Waiting} field to the 
thread body.  The value in \texttt{Waiting} provides a hint to the \texttt{unlock} operation that contention
is present and  waiters exist and handoff can be accomplishedi solely via the \texttt{Grant} field without resorting
to accesses to the lock's \texttt{Tail} field. 
This form also acts to reduce accesses, under contention, to the 
lock's \texttt{Tail} field, reducing coherence traffic.  Races between the threads in \texttt{unlock} path
and new arrivals are possible but benign.  

\lstset{label={Listing:hemlockU}}
\lstset{caption={Hemlock with Optimized Hand-Over -- Variant 3}}
\begin{listing}[htp]    
\begin{adjustwidth}{-0em}{0pt}
\begin{lstlisting}[language=Python,mathescape=true,escapechar=\%]
class Thread :
  atomic<Lock *> Grant = null 

class Lock :
  atomic<Thread *> Tail = null 

def %\underline{Lock}% (Lock * L) :
  assert Self$\rightarrow$Grant $==$ null
  ## constant-time arrival "doorway" step
  auto pred = swap (&L$\rightarrow$Tail, Self) 
  if pred $\neq$ null :
     ## waiting phase 
     while cas(&pred$\rightarrow$Grant, L, null) $\neq$ L : Pause
  
def %\underline{Unlock}% (Lock * L) :  
  assert Self$\rightarrow$Grant $==$ null
  if L$\rightarrow$Tail $\neq$ Self : 
     PassLock: 
     Self$\rightarrow$Grant = L 
     while FetchAdd(&Self$\rightarrow$Grant, 0) $\neq$ null : Pause
     return
  auto v = cas (&L$\rightarrow$Tail, Self, null)
  assert v $\neq$ null 
  if v $\neq$ Self : goto PassLock

\end{lstlisting}
\end{adjustwidth}
\end{listing}

The form in Listing-\ref{Listing:hemlockU} checks for the existence of successors
in the unlock operator by first fetching the lock's \texttt{Tail} field.  Successors exist if and only iff the 
value is not equal to \texttt{Self} (Listing-\ref{Listing:hemlockU} line 12).  
This is tantamount to ``polite'' \texttt{CAS} operator that first loads the value, 
avoiding the futile \texttt{CAS} and its consequent write invalidation when there are successors.  
This form is also immune to use-after-free concerns.  
Under contention, when there are waiting threads, the naive form 
incurs a futile \texttt{CAS} and write invalidation on the \texttt{Tail} field 
(Listing-\ref{Listing:hemlock0} line 16) in the critical path, \emph{before} effecting
transfer at line 20, while this version avoids the futile \texttt{CAS}.  

\section{Waiting Strategies} 

If desired, threads in the Hemlock slow-path (Listing-\ref{Listing:hemlock0} Line 10) 
could optionally be made to wait politely, voluntarily surrending their CPU and blocking 
in the operating system, via constructs such as \texttt{WaitOnAddress}\cite{WaitOnAddress}, 
where a waiting thread could use \texttt{WaitOnAddress} to monitor its 
predecessor's \texttt{Grant} field.

\Invisible{
We note that user-mode locks are not typically implemented as pure spin locks, instead
often using a spin-then-park waiting policy which voluntarily surrenders the CPUs of waiting threads
after a brief optimistic spinning period designed to reduce the context switching rate.
In our case, we find that user-mode is convenient venue for experiments, and note 
that threads in the Hemlock slow-path could easily be made to wait politely
via constructs such as \texttt{WaitOnAddress}\cite{WaitOnAddress},
where a waiting thread could use \texttt{WaitOnAddress} to block, monitoring
its predecessor's \texttt{Grant} field. 
}
 
Under Hemlock, a thread releasing a lock can determine with certainty -- based on the \texttt{Tail} value --
that successors do or do not exist, but the identity of the successor is not known to the thread 
calling \texttt{unlock}.  As such, Hemlock is not immediately amenable to identity-based waiting facilities
such as \texttt{park-unpark} \cite{ParkUnpark,arxiv-Malthusian,LEA2005293} where \texttt{unpark} 
wakes a specific thread. 

To allow purely local spinning and enable the use of \texttt{park}-\texttt{unpark} 
waiting constructs, we can replace the per-thread \texttt{Grant} field with a per-thread pointer 
to a chain of \emph{waiting elements}, each of which represents a waiting thread.
The elements on $T$'s chain are $T$'s immediate successors for various locks.  
Waiting elements contain a \texttt{next} field, a \texttt{flag} and a reference 
to the lock being waited on and can be allocated on-stack. 
Instead of busy waiting on the predecessor's \texttt{Grant} field, waiting 
threads use \code{CAS} to push their element onto the predecessor's chain,
and then busy-wait on the \texttt{flag} in their element. 
The contended unlock($L$) operator detaches the thread's own chain, using 
\texttt{SWAP} of \texttt{null}, traverses the detached chain, and 
sets the \texttt{flag} in the element that references $L$.  
(At most one element will reference $L$).  
Any residual non-matching elements are returned to the chain.  
The detach-and-scan phase repeats until a matching successor is found 
and ownership is transferred.  

\Invisible{In practice, to use park-unpark, you need to know the identity 
of the thread you want to wake up.  In Hemlock, when we release a lock, while we 
can determine with certainty whether a waiter exists, we don’t really know ``who'' is 
waiting, making naive use of park-unpark harder to achieve.  For MCS and CLH, 
for instance, we can encode the ``who'' identity in the queue node, making it 
trivial to use park-unpark.} 

\section{Mitigations for multi-waiting}
As noted above, if a particular lock site is well-balanced, an implementation could opt
to allocate the \texttt{Grant} field for that site on-stack in order to reduce the odds 
of multi-waiting.  

Another simple mitigation is to provision each thread with multiple 
\texttt{Grant} variables, and map the lock address into a per-thread table of such to identify
a specific \texttt{Grant} variable to be used with that particular lock.  The hash function
that maps lock addresses to table indices must be deterministic.   Absent hash collisions,
there can be no multi-waiting.  As multi-waiting is rare and not usually of high degree,
a small table, of, say, 64 elements, suffices.  

\section{HemLock for environments with automatic garbage collection} 
For Java, and similar environments with garbage collection and safe memory reclamation, 
instead of using a per-thread singleton \texttt{Grant} field that resides in 
thread-local storage, the normal approach would be to define a \texttt{WaitElement} structure
that contains the \texttt{Grant} field.  In the \texttt{lock} path, we would first allocate a new 
\texttt{WaitElement} instance with \texttt{new} and then install that reference in the 
\texttt{Tail} field via an atomic exchange.  This approach eliminates the need to clear
the \texttt{Grant} field or wait for the \texttt{Grant} field to become vacant in
the contended \texttt{unlock} path.  The algorithm is no longer \emph{address-based}.  
In addition, we can aggressively store into the element's \texttt{Grant} field in the 
\texttt{unlock} path without concern for use-after-free errors, in anticipation of contended handover.
The downside to this approach is that the address of the element allocated in the \texttt{lock} phase
must be communicated as context to the corresponding \texttt{unlock} operation.  
If necessary, that address can be communicated via a field in the lock body -- protected
by the lock itsef -- or by other means.  In addition, we can 
use simple boolean or \texttt{int} flags for the \texttt{Grant} field, instead
of specifying lock addresses. Listing-\ref{Listing:HemLockJava} provides a sketch
reflecting this approach.  
We note, however, we have come some distance toward effectively recapitulating CLH
in Java, and that a Java version of CLH would be preferrable to and more simple than a
HemLock variant.  

\lstset{label={Listing:HemLockJava}}
\lstset{caption={Hemlock for a Java-like environment}}
\begin{listing}[htp]    
\begin{adjustwidth}{-0em}{0pt}
\begin{lstlisting}[language=Python,mathescape=true,escapechar=\%]
class WaitElement :
  atomic<int> Grant = null 
  atomic<int> Waiting = null 

class Lock :
  atomic<WaitElement *> Tail = null 
  atomic<WaitElement *> Owner = null 

def %\underline{Lock}% (Lock * L) :
  Self = new WaitElement() 
  auto pred = swap (&L$\rightarrow$Tail, Self) 
  if pred $\neq$ null :
     pred$\rightarrow$Waiting = 1 
     while pred$\rightarrow$Grant $==$ 0 : Pause
  L$\rightarrow$Owner $=$ Self 
     
def %\underline{Unlock}% (Lock * L) :  
  Self = L$\rightarrow$Owner  
  Self$\rightarrow$Grant = 1 
  if Self$\rightarrow$Waiting $\neq$ 0 : return 
  cas (&L$\rightarrow$Tail, Self, null)

\end{lstlisting}
\end{adjustwidth}
\end{listing}

\section{Serially Reusable Locking infrastructure} 
Lets say our program can take interrupts or signals, but we will want to use
HemLock in those signal handers.  For simplicity, we assume a locking protocol
where the set of locks accessed in normal non-signal mode is entirely disjoint from the set of
locks that might be accessed within signal handlers.  A thread $T1$ might be running in the
HemLock \texttt{unlock} path and have stored into it's \texttt{Grant} field.  A signal arrives
while \texttt{Grant} is occupied.  In the signal handler, while still running as $T1$, we 
call \texttt{lock} and then \texttt{unlock}, and need to store into $T1$'s \texttt{Grant} to 
effect succession.  By storing into \texttt{Grant}, however,
we have \emph{stomped} and overwriten a value necessary for the interrupted \texttt{unlock} operation, 
which may result in subsequent progress failure when we unwind and return from the signal handler.
As such, we say that HemLock is not a \emph{Serially Reusable Resource}\cite{SeriallyReusable}.  
We observe, however, that if we modify HemLock \texttt{unlock} to save and restore the \texttt{Grant} value, into
an on-frame lock variable, we can easily make HemLock safely serially usable for environments where 
interrupt or signal handling is LIFO.  

We note that MCSH\cite{MCSH} is serially reusable, as is classic MCS in the 
specific situations where the usage is lexically balanced and the MCS waiting \emph{queue node} might 
be allocated on the stack.  
Fissile Locks\cite{dice2020fissile}, using the \emph{Deferred wakeup of the CNA successor}
variation, are also serially reusable.  
Finally, Reciprocating Locks\cite{PPoPP25-dice} are serially reusable when we allocate
the waiting element on the stack instead of thread-local storage\cite{dice2025reciprocatinglocks}
(Appendix B ``On-stack allocation of Wait Elements'').

\end{document}